\documentclass[acmsmall]{acmart}

\AtBeginDocument{%
  }

\setcopyright{acmlicensed}
\copyrightyear{2018}
\acmYear{2018}
\acmDOI{XXXXXXX.XXXXXXX}

\acmJournal{JACM}
\acmVolume{37}
\acmNumber{4}
\acmArticle{111}
\acmMonth{8}

\usepackage{multirow}
\usepackage{anyfontsize}

\usepackage{longtable}

\usepackage{graphicx}
\usepackage{subcaption}

\usepackage{tablefootnote}
\usepackage{amssymb}

\ExplSyntaxOn
\NewDocumentCommand{\longdash}{ O{2} }
 {
  --\prg_replicate:nn { #1 - 1 } { \negthinspace -- }
 }
\ExplSyntaxOff

\newtheorem{mydef}{Definition}
\usepackage{array}

\begin{document}

\title{A Deep Dive into Fairness, Bias, Threats, and Privacy in Recommender Systems: Insights and Future Research}

\author{Falguni Roy}
\authornote{Corresponding author.}
\email{falguniroy@hust.edu.cn}
\email{falguniroy.iit@nstu.edu.cn}
\orcid{0000-0002-0427-5760}
\affiliation{%
  \institution{School of Computer Science and Technology, Huazhong University of Science and Technology}
  \city{Wuhan}
  \country{China}
}
\affiliation{%
  \institution{Noakhali Science and Technology University}
  \city{Noakhali}
  \country{Bangladesh}
}

\author{Xiaofeng Ding}
\orcid{0000-0001-5054-8515}
\affiliation{%
  \institution{School of Computer Science and Technology, Huazhong University of Science and Technology}
  \city{Wuhan}
  \country{China}
  }
\email{xfding@hust.edu.cn}

\author{K.-K. R. Choo}
\orcid{0000-0001-9208-5336}
\affiliation{%
  \institution{Department of Information Systems and Cyber
Security, University of Texas at San Antonio}
  \city{San Antonio}
  \state{TX 782490631}
  \country{USA}
  }
\email{raymond.choo@fulbrightmail.org}

\author{Pan Zhou}
\orcid{0000-0002-8629-4622}
\affiliation{%
  \institution{School of Cyber Science and Engineering, Huazhong University of Science and Technology}
  \city{Wuhan}
  \country{China}
  }
\email{panzhou@hust.edu.cn}


\renewcommand{\shortauthors}{F. Roy et al.}

\begin{abstract}

Recommender systems are essential for personalizing digital experiences on e-commerce sites, streaming services, and social media platforms. While these systems are necessary for modern digital interactions, they face fairness, bias, threats, and privacy challenges. Bias in recommender systems can result in unfair treatment of specific users and item groups, and fairness concerns demand that recommendations be equitable for all users and items. These systems are also vulnerable to various threats that compromise reliability and security. Furthermore, privacy issues arise from the extensive use of personal data, making it crucial to have robust protection mechanisms to safeguard user information. This study explores fairness, bias, threats, and privacy in recommender systems. It examines how algorithmic decisions can unintentionally reinforce biases or marginalize specific user and item groups, emphasizing the need for fair recommendation strategies. The study also looks at the range of threats in the form of attacks that can undermine system integrity and discusses advanced privacy-preserving techniques. By addressing these critical areas, the study highlights current limitations and suggests future research directions to improve recommender systems' robustness, fairness, and privacy. Ultimately, this research aims to help develop more trustworthy and ethical recommender systems that better serve diverse user populations.

\end{abstract}



\received{20 February 2007}
\received[revised]{12 March 2009}
\received[accepted]{5 June 2009}

\maketitle

\section{Introduction}

The \textbf{recommender system (RS)} is becoming an essential part of daily internet usage, offering a one-door solution to the problem of information overload. It retrieves relevant information or services from the vast amount available on the web, tailored to users' needs. RS benefits all internet users, including both service providers and seekers. Its application spans various domains, such as e-commerce websites \cite{ricci2021recommender} like Amazon and Alibaba, entertainment platforms \cite{hasan2019item} like YouTube, TikTok, and Chorki, social networking sites like Facebook, tourism services \cite{hamid2021smart} like Tripadvisor, and lifestyle apps \cite{wan2020addressing} like Yelp and Taobao. Additionally, RS is utilized in healthcare \cite{valdez2019users}, IoT \cite{zhang2022homomorphic}, and numerous other fields, demonstrating its versatility and adaptability across different domains.

A recommender system operates effectively by leveraging both user feedback and an algorithm. User feedback is categorized into implicit and explicit forms \cite{chen2023bias, ogunseyi2023systematic, shi2023selection}. Implicit feedback encompasses behaviours such as a user's browsing history, link clicks, and the frequency with which items are viewed or clicked. This feedback form is generally less precise because it does not involve direct user input and is often interpreted as partial positive feedback. As a result, implicit feedback is not always a reliable indicator of a user's valid preferences \cite{shi2023selection}. Ratings, reviews, and explicit user approval or disapproval expressions characterize explicit feedback. The active engagement of users in providing this feedback makes it a more accurate reflection of their genuine sentiments, thus rendering it more dependable than implicit feedback. Furthermore, explicit feedback can reveal both positive and negative aspects of user preferences. Based on the user feedback, a user-item interaction matrix can be constructed. This matrix includes both explicit and implicit user data. Explicit data represents the historical actions and interaction patterns between users and items, while implicit data pertains to the underlying characteristics of user preferences and demographic information \cite{base01_2021Gender}. However, this implicit data can pose a risk, as adversaries may exploit it to deduce sensitive user information. They can apply inference techniques and machine learning (ML) classification algorithms to the user-item interaction matrix \cite{base01_2021Gender, weinsberg2012blurme}. 

A recommender system (RS) primarily aims to meet service seekers' needs. To achieve this, the RS relies on several standard parameters, referred to as properties, which determine the system's success. The system's success is typically measured by its performance, which directly impacts service seekers' satisfaction. Higher performance is linked to an increased likelihood of system reuse, resulting in economic benefits. These properties help evaluate a recommender approach that should be implemented in a specific domain to fulfil users’ needs through precise recommendations.

The most commonly used performance measurement properties include User Preference \& Fairness, Prediction Accuracy, Coverage, Trust, Diversity, Privacy, and Scalability \cite{ricci2021recommender}. The \textbf{User Preference \& Fairness} ensures that the RS can capture each user’s preferences and treat every user’s preferences equally when generating recommendation lists. \textbf{Prediction Accuracy} is another crucial parameter and the most popular one for measuring performance. The basic assumption is that the more accurately the system predicts users’ preferences, the more likely users will reuse the system. \textbf{Coverage} is another vital parameter for assessing the system’s performance. While prediction accuracy focuses on predicting near-actual ratings for a small subset of items, coverage measures the system’s performance by considering the proportion of users and items it utilizes to meet usage needs. The \textbf{Trust} parameter refers to the system’s credibility, which can be defined in user and item spaces. In the user space, trust indicates users’ confidence in the recommendations and the system’s reliability in safeguarding their data. Users' trust can be enhanced by providing accurate recommendations and ensuring their data is not misused. Trust refers to the system’s confidence in its recommendations in the item space, sometimes called the system’s confidence \cite{hasan2019item}. 

\textbf{Diversity} helps measure the variation in the recommendation list. A system typically contains numerous items in different categories, and diversity quantifies the percentage of varying item groups in the recommendation list compared to the complete set of item groups \cite{castells2021novelty}. A diverse recommendation list should include items similar to the user’s preferences and a mix of popular and non-popular items that might not align with the user’s preferences, thereby reducing the filter bubble effect. The filter bubble refers to a situation where the system continually suggests content that aligns with a user’s previous behaviours, preferences, or opinions, creating a narrow, personalized experience. If the recommendation list includes items unknown to the user’s preferences or knowledge, it is termed the \textbf{\textit{novelty}} of the recommender system \cite{castells2021novelty}. 

\textbf{Privacy} is a growing research focus for recommender systems. The RS often requires detailed information about users’ preferences to generate accurate recommendations, usually based on their historical behaviours. However, this information may contain sensitive user details, such as their tastes and demographic information, posing privacy risks. How effectively the system preserves users’ privacy while providing relevant recommendations is a crucial performance parameter. Lastly, \textbf{Scalability} determines how well a recommender system can handle large volumes of data and how easily it can incorporate new data without compromising performance. 

However, the RS can be categorized into collaborative filtering (CF), content-based filtering, and hybrid RS based on how they process and analyze user \& item data. Among these, CF-based approaches are more widely used \cite{melchiorre2021investigating, shi2023selection, neera2021private, wang2020global, zheng2023federated}. CF-based RS collect users' historical preferences as data and uses this information to predict users' future preferences \cite{ge2021privitem2vec}. After providing recommendations, the system collects new user actions and integrates them with previous data. This recurrent process improves the precision of user recommendations as more information is acquired. Typically, CF-based RS employ an explore-exploit trade-off to execute recommendation tasks. However, this approach can be problematic in terms of fairness and privacy \cite{melchiorre2020personality, melchiorre2021investigating, lambrecht2019algorithmic, neera2021private, wang2020global}. 

When an RS behaves inappropriately towards different groups of users and items, it is considered biased. A biased system generates biased recommendation lists, leading users to interact with these lists (by clicking, rating, or browsing), which further reinforces the system's biased behaviour as these actions are incorporated into the data. If user actions are biased or the system's data is biased, the RS algorithm will produce biased recommendations. Consequently, a biased system fails to treat all user groups and items equally, undermining fairness. Additionally, a biased system cannot ensure item coverage and diversity. Moreover, biased systems are vulnerable to privacy breaches, where adversaries can exploit biased recommendations to access users' private information, posing a threat to the RS \cite{base01_2021Gender, weinsberg2012blurme}. These privacy vulnerabilities and threats affect users' trust and reliability in the system, impacting their satisfaction and the system's economic performance. In conclusion, bias can be introduced at any stage of the RS$—$during data collection, algorithm execution, or user actions$—$negatively affecting the system's fairness. Furthermore, a biased system introduces various threats and privacy vulnerabilities.

\subsection{Existing Surveys and Current Study Contribution}

Given the significant impact and widespread interest in recommendation research, numerous recent surveys have comprehensively reviewed this dynamic field. These surveys can be categorized as either topic-specific, such as session-based RS \cite{wang2021survey}, graph-based RS \cite{wu2022graph, sharma2024survey}, self-supervised leaning-based RS \cite{yu2023self}, reinforcement learning-based RS \cite{afsar2022reinforcement}, or based on performance parameters, focusing on aspects such as accuracy \cite{wu2022survey}, privacy \cite{himeur2022latest, ogunseyi2023systematic, heurix2015taxonomy}, fairness \cite{zehlike2022fairness, jin2023survey}, or bias \cite{chen2023bias} aspects. The main distinction between the current and existing studies lies in considering the recommender system's beyond-accuracy perspectives. This paper thoroughly examines the system's fairness, bias, threat, and privacy aspects. Additionally, this study helps researchers understand these concepts and elucidates the interconnections between these aspects, demonstrating how a violation in one area can compromise another. To the best of our knowledge, this is the first study to showcase the interrelation between fairness, bias, threat, and privacy aspects, along with the taxonomy of each element. 

\subsection{Contributions of This Survey}

The contributions of this survey can be summarized as follows:

\begin{enumerate}

    \item We systematically identify relevant papers on the beyond-accuracy perspectives of recommender systems, focusing on fairness, bias, threat, and privacy. The detailed process of paper collection is mentioned in the section \ref{sec:paper_col}, and Figure \ref{fig:two_images} shows the number of final selected relevant papers (a) by venues \& (b) by years. 

    \begin{figure}[htbp]
    \centering
    \begin{subfigure}[b]{0.52\textwidth}
        \centering
        \includegraphics[width=\textwidth, height=12cm]{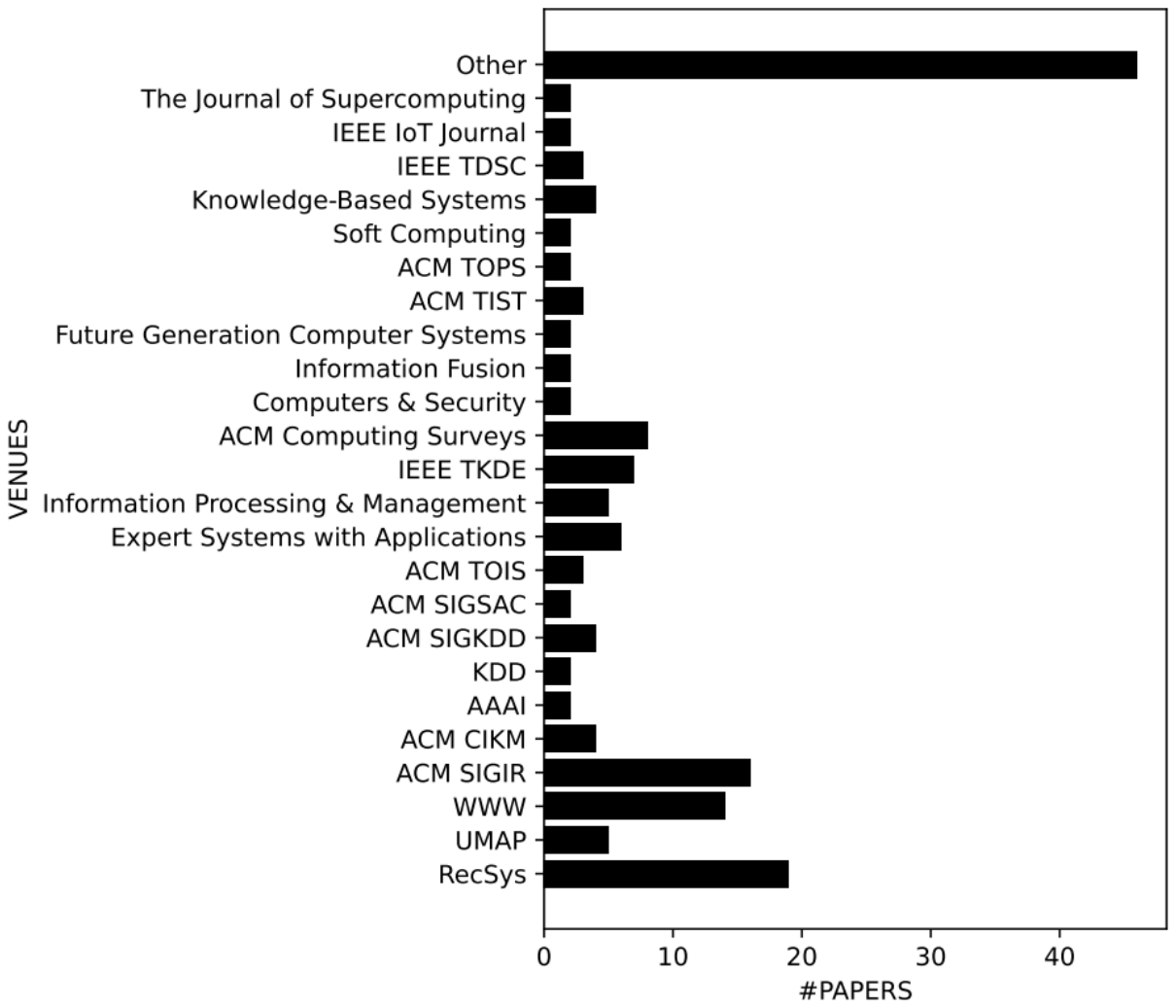} 
        \caption{}
        \label{fig:venue}
    \end{subfigure}
    \hfill
    \begin{subfigure}[b]{0.44\textwidth}
        \centering
        \includegraphics[width=\textwidth, height=4.5cm]{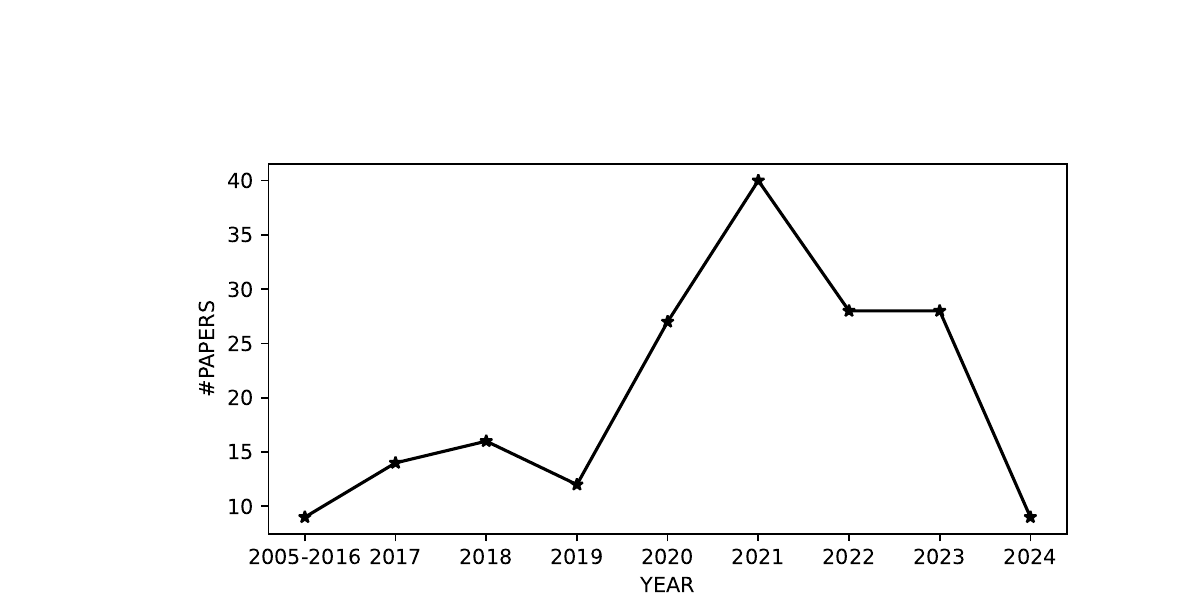} 
        \caption{}
        \label{fig:year}
    \end{subfigure}
    \caption{Distributions of the reviewed papers over venues (a) and over years (b).}
    \label{fig:two_images}
\end{figure}

    \item We present a comprehensive taxonomy for each perspective, enabling researchers to capture the connections and interrelations among these components. The fairness (Figure \ref{fig:FairClassification}) and bias (Figure \ref{fig:BiasClassification}) taxonomy comprises five groups of fairness criteria and twelve groups of bias criteria, respectively. Additionally, threats (Figure \ref{fig:AttackClassification}) are categorized into three groups of attacks for knowledge notation, ten for subject-to-threat notation, and three for intention notation. Furthermore, the privacy perspective encompasses the existing privacy preservation approaches (Figure \ref{fig:pprs}), with four classes for architecture and eleven for methodology notation.

    \item We organize existing literature into summary tables, with each table created according to the relevant viewpoint.

    \item We also point out the one-to-one correlation between fairness and bias with relevant literature (Table \ref{tab:fair} \& \ref{tab:bias}).

    \item We provide a table (Table \ref{tab:threatmodel}) outlining the threat model for each attack, enabling researchers to understand the resources and intentions needed to carry out the attack. This threat model also includes the attack opportunity and potential countermeasures.

    \item We also highlight the advantages and disadvantages of different architectures (Table \ref{tab:ArchiPPRS}) and methodologies (Table \ref{tab:MethodPPRS}) for privacy preservation. This will help researchers understand which types of privacy should be ensured with each architecture and methodology, thereby enhancing the system's overall privacy.

    \item We reference the origins of datasets utilized in privacy research as footnotes. This allows researchers to easily access the datasets for their own studies, using the corresponding paper as a baseline.

    \item Future research directions are highlighted based on our discussion, aiming to attract more scholars to participate in this area of study.
    
\end{enumerate}

\subsection{Paper Collection Methodology}
\label{sec:paper_col}

The survey examines recommender systems focusing on fairness, bias, threats, and privacy. We retrieved publications from leading conferences and journals, including WWW, SIGIR, KDD, AAAI, WSDM, RecSys, CIKM, UMAP, TKDE, TOIS, and Information Processing \& Management to gather relevant papers (Figure \ref{fig:venue}). Additionally, we used Google Scholar as a primary search tool to find recent research using keywords such as \textit{Bias+RS, Fairness+RS, RS+Threat, Privacy+RS, Privacy preservation+RS, collaborative filtering+Privacy}, and similar terms. After collecting the relevant papers, we analyzed citation graphs to identify the most pertinent works from 2017 onward. We also thoroughly reviewed classic and influential papers on recommender systems to ensure we did not miss any crucial work.


\subsection{Scope and Study Structure}

This comprehensive study is expected to offer substantial advantages to four stakeholders within the recommendation community. First, it will be a valuable resource for new researchers and practitioners in recommender systems, helping them quickly familiarize themselves with the field. Second, it will offer existing researchers and practitioners clear guidance on aspects of recommender systems beyond accuracy. Third, the study will be a valuable resource for the inquisitive individual seeking more profound insights. Lastly, for developers currently working on recommender systems, the study will provide practical guidance and insights to help build a fair and error-free system, emphasizing the importance of evaluating performance beyond accuracy.

The survey is organized as follows: Section \ref{sec:fair} addresses fairness issues concerning each actor in the recommender system. Section \ref{sec:bias} classifies biases, including the impact of these biases on fairness within the RS. Section \ref{sec:threat} explores the vulnerabilities of the RS, particularly in the context of various attacks. In contrast, Section \ref{sec:privacy} provides a detailed taxonomy of privacy preservation mechanisms for the RS. Section \ref{sec:fd} outlines future directions to address the limitations identified in this research. Finally, Section \ref{sec:con} presents the main conclusions of the study. For clarity, Table \ref{tab:Abbr} lists this study's most frequently used abbreviations.

\begin{table}[h]
\fontsize{7.5pt}{7.5pt}\selectfont
\renewcommand{\arraystretch}{1.1}
\centering
\caption{Description of Used Abbreviations}
\label{tab:Abbr}
\begin{tabular}{clcl}
\hline
\textbf{Abbreviation} & \multicolumn{1}{c}{\textbf{Description}} & \textbf{Abbreviation} & \multicolumn{1}{c}{\textbf{Description}} \\
\hline
RS & Recommender System & FedRec & Federated Recommender System \\ 
PPRS & Privacy-Preserving Recommender System & SMC & Secure Multiparty Computation \\
AIA & Attribute Inference Attack & MIA & Membership Inference Attack \\
ML & Machine Learning & DP & Differential Privacy \\
HE & Homomorphic Encryption & AL & Adversarial Learning \\
PHE & Partial Homomorphic Encryption & SWHE & Some What Homomorphic Encryption \\
FHE & Fully Homomorphic Encryption & DCS & Distributed Client Server \\
ABE & Attribute-Based Encryption & POI & Point of Interest \\
ML@ & MovieLens Dataset & LFM@ & LastFM Dataset \\
\hline
\end{tabular}
\end{table}

\section{Fairness of Recommender System}
\label{sec:fair}

Fairness is a crucial factor in determining the success of a recommender system (RS) and is closely tied to user preferences. It reflects the system's ability to generate impartial recommendations by treating all groups equally. The RS involves four main actors: consumers, service providers, system owners, and non-participating actors \cite{ricci2021recommender, banerjee2023fairness}. Consumers are the users who receive recommendations, while service providers supply the content or services, such as book publishers \& authors in book recommender systems or artists in music recommender systems. System owners are organizations that manage the platform and recommendation functionalities, with examples including MovieLen, IMDB, eBay, and Tripadvisor. Non-participating actors are those indirectly affected by the RS but not directly involved \cite{banerjee2023fairness}. 

Fairness in an RS ensures equal treatment and fair predictions across all actors \cite{base01_2021Gender, yao2017beyond}. Fairness can be categorized into several types: consumer fairness (c-fairness), service provider fairness (p-fairness), multi-sided fairness (cp-fairness), system owner fairness (o-fairness), and societal fairness (s-fairness) \cite{base01_2021Gender, banerjee2023fairness, burke2018balanced, ekstrand2018privacy, banik2023understanding}. 

\begin{itemize}

    \item \textbf{Consumer fairness (c-fairness)} ensures that recommendations are fair to all user classes, such as those defined by age, gender, occupation, nationality, race, or ethnicity \cite{base01_2021Gender, serbos2017fairness, rahmani2022role}. A breach of c-fairness can lead to population bias.

    \item \textbf{Service provider fairness (p-fairness)} measures fairness among providers by treating all providers' items equally \cite{ferraro2021break, lin2022quantifying, yang2021exploring, gomez2022provider}. Violations of p-fairness can result in item-oriented biases, either from over-representation of specific preferences or unfair treatment of providers to optimize system performance \cite{banerjee2023fairness}. P-fairness can be managed through post-processing techniques \cite{gomez2022provider}.

    \item \textbf{Multi-sided fairness (cp-fairness)} ensures that both users and service providers are treated fairly in recommendations \cite{base01_2021Gender, burke2018balanced}.

    \item \textbf{System owner fairness (o-fairness)} assesses the fairness of the platform owner towards other actors \cite{banik2023understanding, banerjee2023fairness, gupta2022fairfoody}. A lack of o-fairness can lead to position and inductive biases \cite{zhu2020measuring, gupta2021online, kokkodis2020your, mavridis2020beyond, yang2021local}.

    \item \textbf{Societal fairness (s-fairness)} focuses on non-participating actors who are indirectly impacted by the RS, such as local businesses and political entities.  The s-fairness mainly implies the tourism recommender system. A lack of it can raise sustainability issues \cite{merinov2023sustainability, pachot2021multiobjective, patro2020towards}.
    
\end{itemize}

\begin{figure}[h]
    \centering
    \includegraphics[width=.95\linewidth]{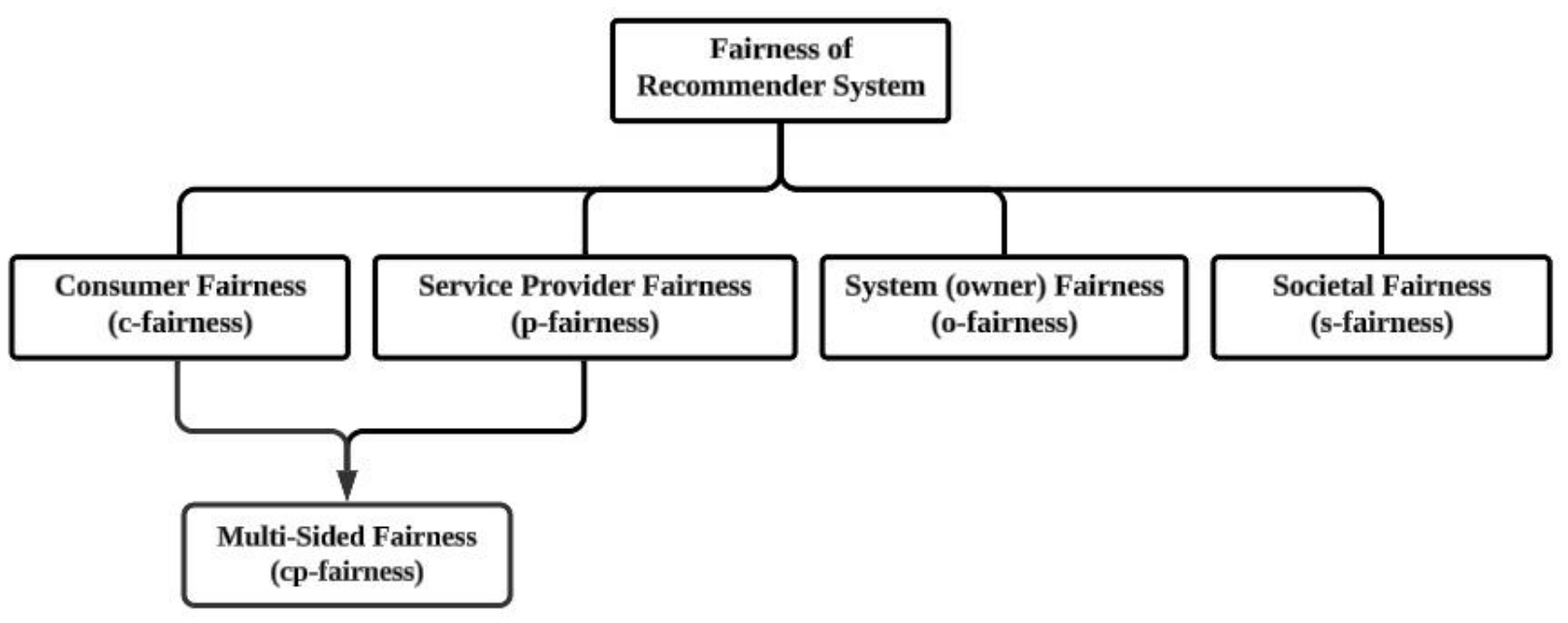}
    \caption{Classification of Fairness of Recommender system}
    \label{fig:FairClassification}
\end{figure}

Fairness in an RS can be evaluated through the ranking of the recommended items list \cite{zehlike2020reducing}, the prediction accuracy of the recommendation \cite{wan2020addressing, yao2017beyond, beutel2017beyond}, and pairwise metrics \cite{misztal2023bias}. When an RS lacks or violates fairness, it can lead to biased treatment of actors, reducing system performance and user satisfaction. Thus, fairness is integral to system performance and user satisfaction. Figure \ref{fig:FairClassification} illustrates the classification of recommender system fairness, and Table \ref{tab:fair} details each fairness type and its potential violations, supported by reviewed publications.

\begin{table}[h]
\fontsize{7.5pt}{7.5pt}\selectfont
\renewcommand{\arraystretch}{1.1}
\centering
\caption{A Lookup Table of Fairness Criteria and Associated Bias of the Recommender System}
\label{tab:fair}
\begin{tabular}{>{\centering\hspace{0pt}}m{0.171\linewidth}>{\hspace{0pt}}m{0.472\linewidth}>{\centering\hspace{0pt}}m{0.156\linewidth}>{\centering\arraybackslash\hspace{0pt}}m{0.112\linewidth}}
\hline
\textbf{Fairness Type}& \multicolumn{1}{>{\centering\hspace{0pt}}m{0.472\linewidth}}{\textbf{Focused Criteria}} & \textbf{Consequence of} \par{}\textbf{Fairness Violation} & \textbf{Reviewed}\par{}\textbf{Publications} \\ 
\hline

Consumer Fairness\par{}(c-fairness) & Ensure equal treatment of all system users who are supposed to receive the recommendations (individual and group fairness) & Population Bias & \cite{base01_2021Gender, ekstrand2018privacy, serbos2017fairness, rahmani2022role}  \\

Service Provider Fairness (p-fairness) & Verify the equal treatment of all service providers who are supposed to provide their service for recommendations & Popularity Bias, Exposure Bias, Homogeneity Bias, Media Bias, Selection Bias, Positive Bias, Position Bias \& Conformity Bias & \cite{ferraro2021break, lin2022quantifying, yang2021exploring, gomez2022provider, zehlike2020reducing} \\

Multi-sided Fairness (cp-fairness) & Ensure equal treatment of all system users and service providers for the recommendations & Data \& Algorithmic Bias & \cite{burke2018balanced} \\ \\

Platform Owner Fairness (o-fairness) & Define equal treatment behaviour of the recommendation platform towards all system users and service providers & Algorithmic Bias & \cite{gupta2022fairfoody, zhu2020measuring, gupta2021online, kokkodis2020your, mavridis2020beyond, yang2021local} \\

Societal Fairness (s-fairness) & Measure the fairness of the non-participating actors who are affected by the RS but not directly associated with it & Sustainability Issues \& Media Bias &  \cite{banerjee2023fairness, banik2023understanding, merinov2023sustainability, pachot2021multiobjective, patro2020towards} \\

\hline
\end{tabular}
\end{table}

\section{Bias in Recommender System}
\label{sec:bias}

According to the general definition, "Bias is a strong feeling in favour of or against one group of people, or one side in an argument, often not based on the fair judgment" \cite{breitfuss2021representing}. In a recommender system, bias can be defined as the "uneven treatment of different groups of the same object where an object can be either user or item of the system". The bias often arises because the system prioritizes accuracy over fairness, favouring well-known items for recommendations without considering the satisfaction of all users \cite{ferraro2021break}. The presence of bias in the system highlights the need for fairness in RS and the importance of ensuring equitable treatment for all users. Typically, bias in RS is observational rather than experimental, as it collects user interaction data to execute simulations within the system \cite{chen2023bias, misztal2023bias}. 

\begin{figure}[ht]
    \centering
    \includegraphics[width=0.9\linewidth]{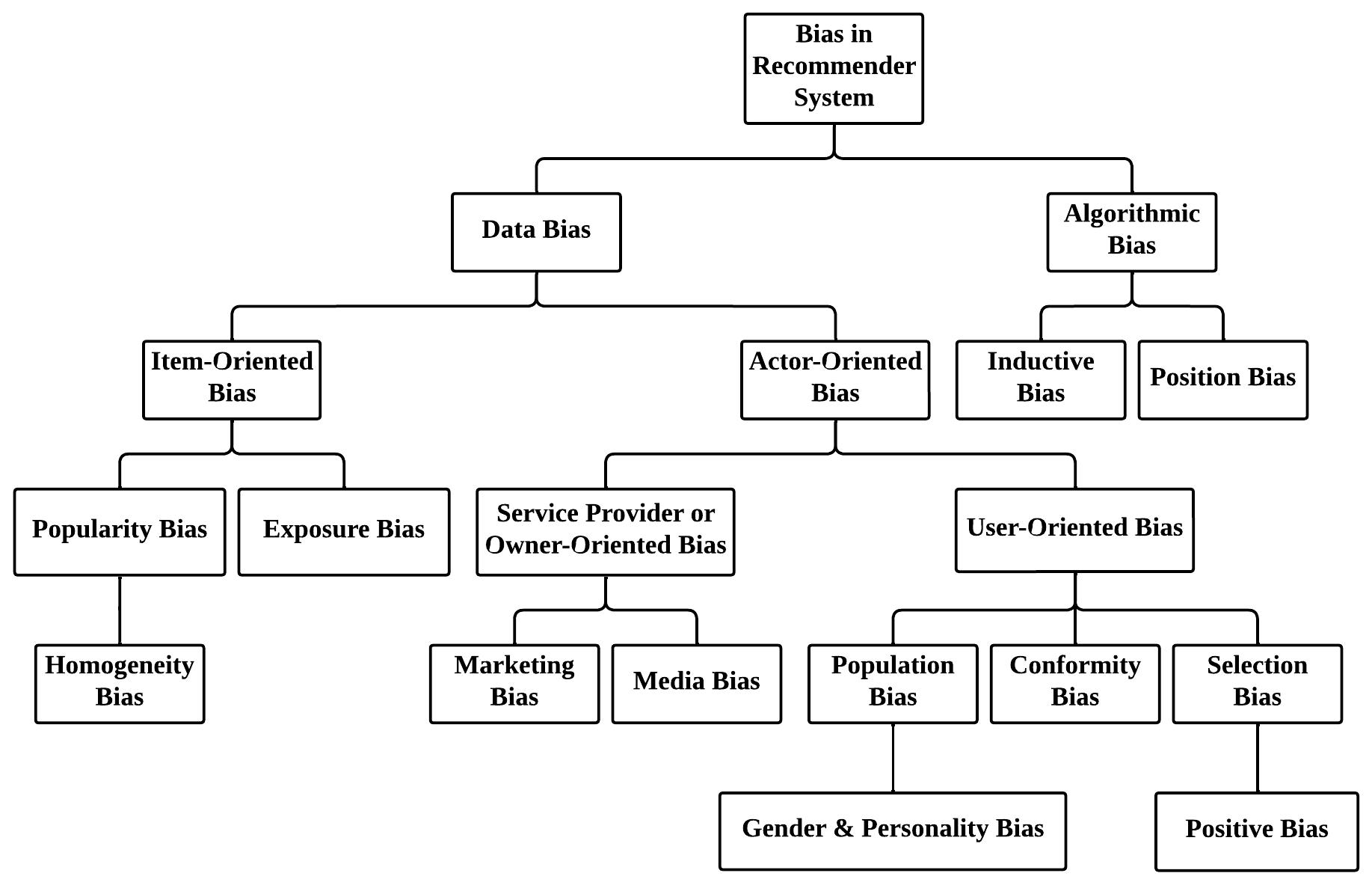}
    \caption{Taxonomy of Bias in Recommender system}
    \label{fig:BiasClassification}
\end{figure}

Bias in RS can be categorized based on its components into data bias and algorithmic bias. Algorithmic bias can be further divided into inductive bias and position bias. Additionally, based on object characteristics, bias in RS can be classified into user-oriented bias and item-oriented bias. These two categories of biases create a feedback loop around the users, resulting in a phenomenon known as "$Filter\; Bubbles$" or "$Echo\; Chambers$". This feedback loop operates according to the Matthew effect, meaning "the rich get richer" \cite{chen2023bias}. Depending on the nature of their origin, user-oriented and item-oriented biases can be subdivided into several other forms. Figure \ref{fig:BiasClassification} provides a detailed classification of bias in RS.

A recommender system typically operates through a cyclical process, where bias can be introduced at any stage (data collection, recommendation generation, or users' actions )and can exacerbate existing biases. There are three main reasons for introducing and persisting bias in the system. First, users exhibit imbalanced behaviour toward different groups of users and items. Second, there is an uneven representation of items in the dataset. Third, the system's feedback loop characteristic contributes to bias. Bias in the system reduces diversity and user satisfaction, negatively affecting the service provider's long-term revenue. Also, the magnitude of bias depends on the data size, and bias can be found in large proportion when the data size is scanty. The relation between bias \& data size can be formulated as follows:

\begin{equation}
  bias \propto \frac{1}{datasize}
\end{equation}

Bias in recommendation systems can manifest in various forms, depending on how it is created, and all forms of bias are detailed in the following section.

\begin{enumerate}
    \item \textbf{Data Bias.} It occurs when the data has varying distributions among different groups. The formal definition of the data bias is shown in Equation \ref{eqn:databias}.

    \begin{equation}
    \label{eqn:databias}
        D = d_{ma}\cup d_{mi} \;\;\;\;\; \text{where}~ d_{ma},d_{mi} \in G \;\;\; \& \;\;\; d_{ma} >  d_{mi}
    \end{equation}

    Here, $D$ is the dataset \& $G$ is the data group that can be distinguished by users' gender, age, country, or items' popularity. $d_{ma}$ \& $d_{mi}$ denote the distribution major and minor data group respectively. 

    \item \textbf{Marketing Bias:} This bias often occurs in product recommender systems, such as those on e-commerce websites. Marketing bias can be introduced by the recommendation platform to achieve economic goals or by under-representing specific market segments. This bias affects the fairness of consumers and service providers, leading to market loss, unsatisfactory user experiences, and other adverse outcomes \cite{wan2020addressing}. 

    \item \textbf{Media Bias:} This bias primarily originates in the news recommender system and can be intentionally or unintentionally introduced by the service provider. Media bias is the tendency of media to present news in a way that aligns with specific interests and beliefs \cite{ruan2023influence}. This bias can result in several disadvantages, including incomplete news coverage, discrimination, and social inequality \cite{ruan2023influence}.  

    \item \textbf{Population Bias: } In recommender systems, population bias refers to the distortion in the data of a given population compared to the target population \cite{chizari2022comparative}. In other words, population bias occurs when the system behaves differently towards its users or service providers based on their demographic information \cite{melchiorre2021investigating}. This bias, also known as \textbf{demographic bias}, can manifest in various forms, such as \textbf{gender bias} \cite{base01_2021Gender, lambrecht2019algorithmic, melchiorre2021investigating, ekstrand2018exploring, ekstrand2018all} and \textbf{personality bias} \cite{melchiorre2020personality}.
       
    \item \textbf{Selection Bias:} The users introduce selection bias through explicit feedback on items they are either highly satisfied or dissatisfied with \cite{shi2023selection}. Due to the large volume of data in the system, every user cannot access every piece of data. Additionally, because users tend to give explicit feedback only on items they feel strongly about, either positively or negatively, the user-item interaction matrix becomes sparse, and the missing values in the matrix are not random \cite{shi2023selection}. This scenario is called the Missing Not At Random (MNAR) issue \cite{chen2018investigating}. The MNAR issue can be addressed by treating all missing values as negative data points, known as the All Missing As Negative (AMAN) approach, or by ignoring all missing values when executing an algorithm, known as the All Missing As Unknown (AMAU) approach \cite{yang2021exploring}. The tendency of users to provide feedback on a select set of items, determining good or bad items, is called Selection Bias. Furthermore, when users more frequently provide feedback on highly satisfied items, it is known as \textbf{Positive Bias} \cite{huang2020keeping, sigir2024accept, park2018positivity}, resulting in an over-representation of positive feedback.
       
    \item \textbf{Conformity Bias:} Users' behaviours also contribute to this bias. Conformity bias arises in the system when users tend to act in line with other users in the group, disregarding their genuine preferences. This bias is frequently observed in users' ratings \cite{lederrey2018sheep, pan2024improved} and reviews \cite{himeur2022latest}. Social factors heavily influence it, as users often mimic their friends' behaviours \cite{wang2017learning, chaney2015probabilistic} or suppress negative opinions after seeing the prevalence of positive public comments \cite{chen2023bias}. Users' implicit feedback can also be influenced by conformity bias; for example, a user might click on an item simply because many other users have done so \cite{zheng2021disentangling}. Conformity bias is also known as \textbf{Sequential Bias} \cite{himeur2022latest}. This bias pressures users to display their genuine preferences toward certain items.
     
    \item \textbf{Popularity Bias:} Popularity bias occurs when the system frequently recommends popular items, while less popular but potentially more relevant items receive little exposure \cite{abdollahpouri2017controlling, rhee2022countering, wei2021model, klimashevskaia2023addressing}. This issue is akin to the class imbalance in machine learning. Popularity bias results in long-tail problems and undermines service providers' fairness (p-fairness) \cite{abdollahpouri2019unfairness, abdollahpouri2020connection}. When an RS recommends items from a limited set of sources due to their popularity or business emphasis, this bias is known as \textbf{Homogeneity Bias} \cite{nikolov2019quantifying, bhadani2021biases}. Popularity bias in the input data has the potential to mislead users away from their actual preferences and trigger position bias. This reduces the level of personalized recommendations and amplifies the Matthew effect, where popular items become even more popular due to increased exposure \cite{chen2023bias, zhu2021popularity}. Additionally, popularity bias can degrade the system's performance \cite{abdollahpouri2019unfairness, mansoury2020investigating}.
        
    \item \textbf{Exposure Bias:} In a recommender system (RS), item visibility and exposure concepts hold different meanings. Item visibility refers to the frequency of an item's appearance in the recommendation ranking \cite{mansoury2021graph}. In contrast, item exposure refers to evaluating an item's position within the ranking \cite{biega2018equity, zehlike2020reducing}. Exposure bias arises when users are shown only a specific subset of items, which may result from the system's policy \cite{liu2020general}, users' active searching and selection criteria \cite{ovaisi2020correcting}, or existing popularity bias within the data \cite{zheng2021disentangling, mansoury2022understanding}. Implicit feedback can contribute to exposure bias since it comprises partial positive interaction data between users and items \cite{chen2023bias}. Moreover, implicit feedback does not account for missing data, which could either indicate negative user preferences or user unawareness of certain items. Consequently, when relying on implicit feedback, the RS tends to recommend items derived from this feedback, leading to exposure bias and compromising the fairness of the RS.  
    
    \item \textbf{Algorithmic Bias:} Algorithmic Bias in recommender systems can arise from an uneven distribution in the dataset on which the algorithm is trained or deliberately introduced during analysis to reflect economic goals or achieve desirable characteristics in the output. Even after eliminating all bias from the data, an algorithm's results may still vary across different data groups due to algorithmic bias \cite{lambrecht2019algorithmic, wang2021user}. This phenomenon is also known as \textbf{Model Bias} \cite{rhee2022countering, melchiorre2021investigating}. Algorithmic bias can be categorized into inductive bias and position bias. \textbf{Inductive Bias} occurs when a bias is intentionally introduced into a model to enhance its performance by better learning the target function for recommendations \cite{chen2023bias, yang2021local, lin2021mitigating}. The output of an RS typically consists of an ordered recommendation list that combines user preferences with item popularity. Users are more likely to interact with items at the top of the recommendation list because these items quickly capture attention \cite{biega2018equity}. Therefore, the order of the recommendation list is crucial, as there is a high probability of users interacting with top-ranked items, regardless of the items' relevance to their needs. This tendency to interact with top-positioned items without considering their relevance creates \textbf{Position Bias} \cite{chen2023bias, biega2018equity, joachims2017accurately, agarwal2019general}. Popularity bias exacerbates position bias, leading to long-tail problems in the system. Consequently, position bias also denotes \textbf{Ranking Bias} \cite{banerjee2023fairness}.

\end{enumerate}

However, the bias in the RS is not necessarily detrimental and can sometimes enhance personalization, leading to greater user satisfaction. The bias in the system is tolerable as long as it does not compromise fairness. When fairness is not maintained, it can undermine the system's diversity and promote user homogenization. Usually, a biased system generates the biased output, and the uncontrolled biases can lead to the uneven distribution of opportunities and resources \cite{biega2018equity, chen2018investigating}, exacerbating the "$Echo\; Chamber$" or "$Filter\; Bubbles$" effects \cite{ge2020understanding, jiang2019degenerate}. Additionally, demographic bias can expose privacy vulnerabilities, allowing adversaries to infer and access users' sensitive information. Table \ref{tab:bias} summarises each type of bias, the associated fairness violations, and relevant literature.
 
{\fontsize{7.5pt}{7.5pt}\selectfont
\renewcommand{\arraystretch}{0.9}
\begin{longtable}{cccllcc}
\caption{A Summary of Existing Biases in the Recommender System} \\
\hline 
\textbf{Bias Type} & \textbf{Subset of} & \textbf{Begets From} & \multicolumn{1}{c}{\textbf{Origination Cause}} & \multicolumn{1}{c}{\textbf{\begin{tabular}[c]{@{}c@{}}Bias Detection\\ Approaches\end{tabular}}} & \textbf{\begin{tabular}[c]{@{}c@{}}Fairness\\ Violation\end{tabular}} & \textbf{\begin{tabular}[c]{@{}c@{}}Reviewed \\ Publications\end{tabular}} \\ 
\hline  \\
\endfirsthead
\multicolumn{6}{c}%
{\tablename\ \thetable\ . A Summary of Existing Biases in the Recommender System (Cont.)} \\ \\
\hline

\textbf{Bias Type} & \textbf{Subset of} & \textbf{Begets From} & \multicolumn{1}{c}{\textbf{Origination Cause}} & \multicolumn{1}{c}{\textbf{\begin{tabular}[c]{@{}c@{}}Bias Detection\\ Approaches\end{tabular}}} & \textbf{\begin{tabular}[c]{@{}c@{}}Fairness\\ Violation\end{tabular}} & \textbf{\begin{tabular}[c]{@{}c@{}}Reviewed \\ Publications\end{tabular}} \\ 
\hline \\ 
\endhead
\hline 
\endfoot
\hline
\endlastfoot

Data Bias & \textbf{\longdash[3]} & \begin{tabular}[c]{@{}c@{}}Algorithmic \\ Bias and/or\\ Imbalance \\ Data\end{tabular} & \begin{tabular}[c]{@{}l@{}}1. For the presence \\ of imbalance group \\ of users/items data,\\ 2. Model induced bias\end{tabular} & 
\begin{tabular}[c]{@{}l@{}}\begin{tabular}{@{\labelitemi\hspace{\dimexpr\labelsep+0.5\tabcolsep}}l@{}}Kullback–Leibler \end{tabular}\\divergence\end{tabular} 
& cp-fairness & \cite{chen2023bias} \\ \\

\begin{tabular}[c]{@{}c@{}}Marketing \\ Bias\end{tabular} & \begin{tabular}[c]{@{}c@{}}Data Bias $\to$ \\ Service \\ Provider\\ Oriented Bias\end{tabular} & \textbf{\longdash[3]} & \begin{tabular}[c]{@{}l@{}}1. under-representation \\ of the market segments,\\ 2. System induced to \\ gain economic goals\end{tabular} & 
\begin{tabular}[c]{@{}l@{}}\begin{tabular}{@{\labelitemi\hspace{\dimexpr\labelsep+0.5\tabcolsep}}l@{}}Statistical analy-\end{tabular}\\ sis between product \\ selection and \\ satisfaction \end{tabular} 
& cp-fairness & \cite{wan2020addressing} \\ \\

\begin{tabular}[c]{@{}c@{}}Media \\ Bias\end{tabular} & \begin{tabular}[c]{@{}c@{}}Data Bias $\to$ \\ Service Provider\\ Oriented Bias\end{tabular} & \textbf{\longdash[3]} & \begin{tabular}[c]{@{}l@{}}1. To emphasize and \\ enforce specific needs,\\ 2. To achieve economic \\ goals by attracting more \\ users\end{tabular} & 
\begin{tabular}[c]{@{}l@{}}\begin{tabular}{@{\labelitemi\hspace{\dimexpr\labelsep+0.5\tabcolsep}}l@{}}Applied data \end{tabular}\\ fusion strategy \\ for distant \\ supervision \\ annotated data \\ to detect bias\end{tabular} 
& \begin{tabular}[c]{@{}c@{}}p-fairness\\ and\\ s-fairness\end{tabular} & \cite{ruan2023influence} \\ 

\begin{tabular}[c]{@{}c@{}}Population \\ Bias\end{tabular} & \begin{tabular}[c]{@{}c@{}}Data Bias $\to$ \\ User-Oriented\\ Bias\end{tabular} & \textbf{\longdash[3]} & \begin{tabular}[c]{@{}l@{}}1. Under-representation \\ of several groups based \\ on demographic in the \\ data,\\ 2. Prioritize active users \\ data for providing \\ recommendation\end{tabular} & 
\begin{tabular}[c]{@{}l@{}}\begin{tabular}{@{\labelitemi\hspace{\dimexpr\labelsep+0.5\tabcolsep}}l@{}}Performance \end{tabular}\\ deviation analysis \\ between group \end{tabular} 
& cp-fairness & \cite{melchiorre2021investigating, chizari2022comparative}  \\  \\

\begin{tabular}[c]{@{}c@{}}Gender, Age,\\ Culture and\\ Personality \\ Bias\end{tabular} & \begin{tabular}[c]{@{}c@{}}Data Bias $\to$ \\ User-Oriented\\ Bias\end{tabular} & \begin{tabular}[c]{@{}c@{}}Population \\ Bias\end{tabular} & \begin{tabular}[c]{@{}l@{}}Under representation of \\ several groups based on \\ demographic\end{tabular} & 
\begin{tabular}[c]{@{}l@{}}\begin{tabular}{@{\labelitemi\hspace{\dimexpr\labelsep+0.5\tabcolsep}}l@{}}Performance \end{tabular}\\ (accuracy and \\ ranked list) \\ deviation analysis \\ between group \end{tabular} 
& cp-fairness & \begin{tabular}[c]{@{}l@{}} \cite{melchiorre2020personality, lambrecht2019algorithmic} \\ \cite{base01_2021Gender, ekstrand2018exploring} \\ \cite{ekstrand2018all} \end{tabular} \\ \\

\begin{tabular}[c]{@{}c@{}}Selection \\ Bias\end{tabular} & \begin{tabular}[c]{@{}c@{}}Data Bias $\to$ \\ User-Oriented\\ Bias\end{tabular} & \textbf{\longdash[3]} & \begin{tabular}[c]{@{}l@{}}Over representation of \\ users' preferences\end{tabular} & \begin{tabular}[c]{@{}l@{}}\begin{tabular}{@{\labelitemi\hspace{\dimexpr\labelsep+0.5\tabcolsep}}l@{}}Skewed ratings ~\end{tabular}\\~ ~ distribution towa- \\~ ~rds set of items\end{tabular} & p-fairness & \begin{tabular}[c]{@{}c@{}} \cite{shi2023selection, chen2018investigating} \\ \cite{yang2021exploring} \end{tabular} \\ \\

\begin{tabular}[c]{@{}c@{}}Positive \\ Bias\end{tabular} & \begin{tabular}[c]{@{}c@{}}Data Bias $\to$ \\ User-Oriented\\ Bias\end{tabular} & \begin{tabular}[c]{@{}c@{}}Selection \\ Bias\end{tabular} & \begin{tabular}[c]{@{}l@{}}Users self positive \\ selection\end{tabular} & \begin{tabular}[c]{@{}l@{}}\begin{tabular}{@{\labelitemi\hspace{\dimexpr\labelsep+0.5\tabcolsep}}l@{}}Analyzing distri-\end{tabular}\\~ ~ bution of ratings\\
\begin{tabular}{@{\labelitemi\hspace{\dimexpr\labelsep+0.5\tabcolsep}}l@{}}Observing the~\end{tabular}\\~ ~independence of \\~ ~the observation \\~ ~ process from \\~ ~ the value of\\~ ~ unobserved data\end{tabular} & p-fairness & \begin{tabular}[c]{@{}c@{}} \cite{huang2020keeping, sigir2024accept} \\ \cite{park2018positivity} \end{tabular} \\ 

\begin{tabular}[c]{@{}c@{}}Conformity \\ Bias, or\\ Sequential \\ Bias\end{tabular} & \begin{tabular}[c]{@{}c@{}}Data Bias $\to$ \\ User-Oriented\\ Bias\end{tabular} & \textbf{\longdash[3]} & \begin{tabular}[c]{@{}l@{}}Obedience to others\\ opinion\end{tabular} & \begin{tabular}[c]{@{}l@{}}\begin{tabular}{@{\labelitemi\hspace{\dimexpr\labelsep+0.5\tabcolsep}}l@{}}Analyzing user- ~\end{tabular}\\~ ~ item affinity from \\~ ~  interaction labels \end{tabular} & p-fairness & \begin{tabular}[c]{@{}c@{}} \cite{lederrey2018sheep, pan2024improved} \\ \cite{himeur2022latest, zheng2021disentangling} \\ \cite{wang2017learning, chaney2015probabilistic} 
\end{tabular} \\ 

\begin{tabular}[c]{@{}c@{}}Popularity \\ Bias\end{tabular} & \begin{tabular}[c]{@{}c@{}}Data Bias $\to$ \\ Item-Oriented\\ Bias\end{tabular} & \textbf{\longdash[3]} & \begin{tabular}[c]{@{}l@{}}Algorithm enforced\\ and imbalance item \\ groups\end{tabular} & \begin{tabular}{@{\labelitemi\hspace{\dimexpr\labelsep+0.5\tabcolsep}}l@{}}Item Coverage\\Gini Index\\Item Entropy\\Average Diversity\\Average Popularity\end{tabular} & p-fairness & \begin{tabular}[c]{@{}c@{}} \cite{abdollahpouri2017controlling, rhee2022countering} \\ \cite{wei2021model, klimashevskaia2023addressing} \\ \cite{abdollahpouri2019unfairness, abdollahpouri2020connection} \\ \cite{zhu2021popularity, mansoury2020investigating} \end{tabular} \\ \\

\begin{tabular}[c]{@{}c@{}}Homogeneity \\ Bias\end{tabular} & \begin{tabular}[c]{@{}c@{}}Data Bias $\to$ \\ Item-Oriented\\ Bias\end{tabular} & \begin{tabular}[c]{@{}c@{}}Popularity \\ Bias\end{tabular} & Algorithm enforced & \begin{tabular}{@{\labelitemi\hspace{\dimexpr\labelsep+0.5\tabcolsep}}l@{}}Item Coverage\\Gini Index\\Item Entropy\\Average Diversity\\Average Popularity\end{tabular} & p-fairness & \cite{nikolov2019quantifying, bhadani2021biases} \\ \\

\begin{tabular}[c]{@{}c@{}}Exposure \\ Bias\end{tabular} & \begin{tabular}[c]{@{}c@{}}Data Bias $\to$ \\ Item-Oriented\\ Bias\end{tabular} & \textbf{\longdash[3]} & \begin{tabular}[c]{@{}l@{}}1. Users response; \\ 2. Items popularity;\\ 3. Algorithm enforced\end{tabular} & \begin{tabular}[c]{@{}l@{}}\begin{tabular}{@{\labelitemi\hspace{\dimexpr\labelsep+0.5\tabcolsep}}l@{}}Visibility Shift in~\end{tabular}\\~ ~ Recommendation \\~ ~ List\\ 
\begin{tabular}{@{\labelitemi\hspace{\dimexpr\labelsep+0.5\tabcolsep}}l@{}}Item Coverage  \\ Gini Index \\ Entropy Check \\ Average Diversity\end{tabular}\end{tabular} & p-fairness & \begin{tabular}[c]{@{}c@{}} \cite{mansoury2022understanding, mansoury2021graph} \\ \cite{biega2018equity, zehlike2020reducing} \\ \cite{liu2020general, ovaisi2020correcting}  
\end{tabular} \\ \\

\begin{tabular}[c]{@{}c@{}}Algorithmic \\ Bias, or \\ Model Bias\end{tabular} & \textbf{\longdash[3]} & \begin{tabular}[c]{@{}c@{}}Data Bias, \\ or System \\ Injected\end{tabular} & \begin{tabular}[c]{@{}l@{}}1. Processing the bias \\ data;\\ 2. For gain economic \\ goals\end{tabular} & 
\begin{tabular}[c]{@{}l@{}}\begin{tabular}{@{\labelitemi\hspace{\dimexpr\labelsep+0.5\tabcolsep}}l@{}}Serendipity Check\\Average Diversity\\Long-Tail Coverage\\Unexpectedness\end{tabular}\\ Check\\\end{tabular} 
& \begin{tabular}[c]{@{}c@{}}cp-fairness\\ and\\ o-fairness\end{tabular} & \cite{lambrecht2019algorithmic, wang2021user} \\ 

\begin{tabular}[c]{@{}c@{}}Inductive \\ Bias\end{tabular} & Algorithmic Bias & \textbf{\longdash[3]} & \begin{tabular}[c]{@{}l@{}}System injects the \\ bias to gain better \\ performance for \\ economic and user \\ satisfaction enhan-\\ cement prespective\end{tabular} & \multicolumn{1}{c}{\textbf{\longdash[3]}} &o-fairness & \cite{yang2021local, lin2021mitigating} \\ \\

\begin{tabular}[c]{@{}c@{}}Position Bias, \\ or Ranking \\ Bias \end{tabular} & Algorithmic Bias & \textbf{\longdash[3]} & \begin{tabular}[c]{@{}l@{}}1. Inclination of inter-\\ act with top-ordered\\ listed items;\\ 2. Over-representation \\ of top lists\end{tabular} & 
\begin{tabular}[c]{@{}l@{}}\begin{tabular}{@{\labelitemi\hspace{\dimexpr\labelsep+0.5\tabcolsep}}l@{}}True Relevancy \end{tabular}\\  Check\\
\begin{tabular}{@{\labelitemi\hspace{\dimexpr\labelsep+0.5\tabcolsep}}l@{}}Propensity Score \end{tabular}\\ Check\end{tabular} 
& \begin{tabular}[c]{@{}c@{}}o-fairness\\ and \\ p-fairness\end{tabular} & \begin{tabular}[c]{@{}c@{}} \cite{biega2018equity, joachims2017accurately} \\ \cite{agarwal2019general, banerjee2023fairness}  \\ 

\end{tabular}
\label{tab:bias}
\end{longtable}
}

\section{Threats of Recommender System}
\label{sec:threat}

As the recommender system gathers users' historical interaction data to determine the relevance of their preferences through analysis, this open and interactive process has both positive and negative aspects. On the positive side, the more data the system collects about users, the better it can understand their preferences and generate accurate recommendations. On the negative side, however, the system becomes susceptible to various threats that can lead to different attacks, especially since the system has economic objectives as well. 

In the context of an RS, an attack refers to any intentional and malicious action or strategy aimed at compromising the system's integrity, accuracy, fairness, or security. These attacks exploit system algorithms, data inputs, or user interaction vulnerabilities to achieve specific objectives. These objectives include promoting or demoting certain items, influencing user behaviour, manipulating public opinion, extracting sensitive information, or rendering the system ineffective. The nature of attacks on an RS can vary depending on the adversary’s intentions and knowledge. Generally, an adversary conducts a user- or item-oriented attack based on their knowledge and perspective. Usually, an adversary could be any party with unlawful motives, such as advertisers, banks, insurance companies, or cyber criminals \cite{jia2017attriinfer}.

\begin{figure}[ht]
    \centering
    \includegraphics[width=0.95\linewidth]{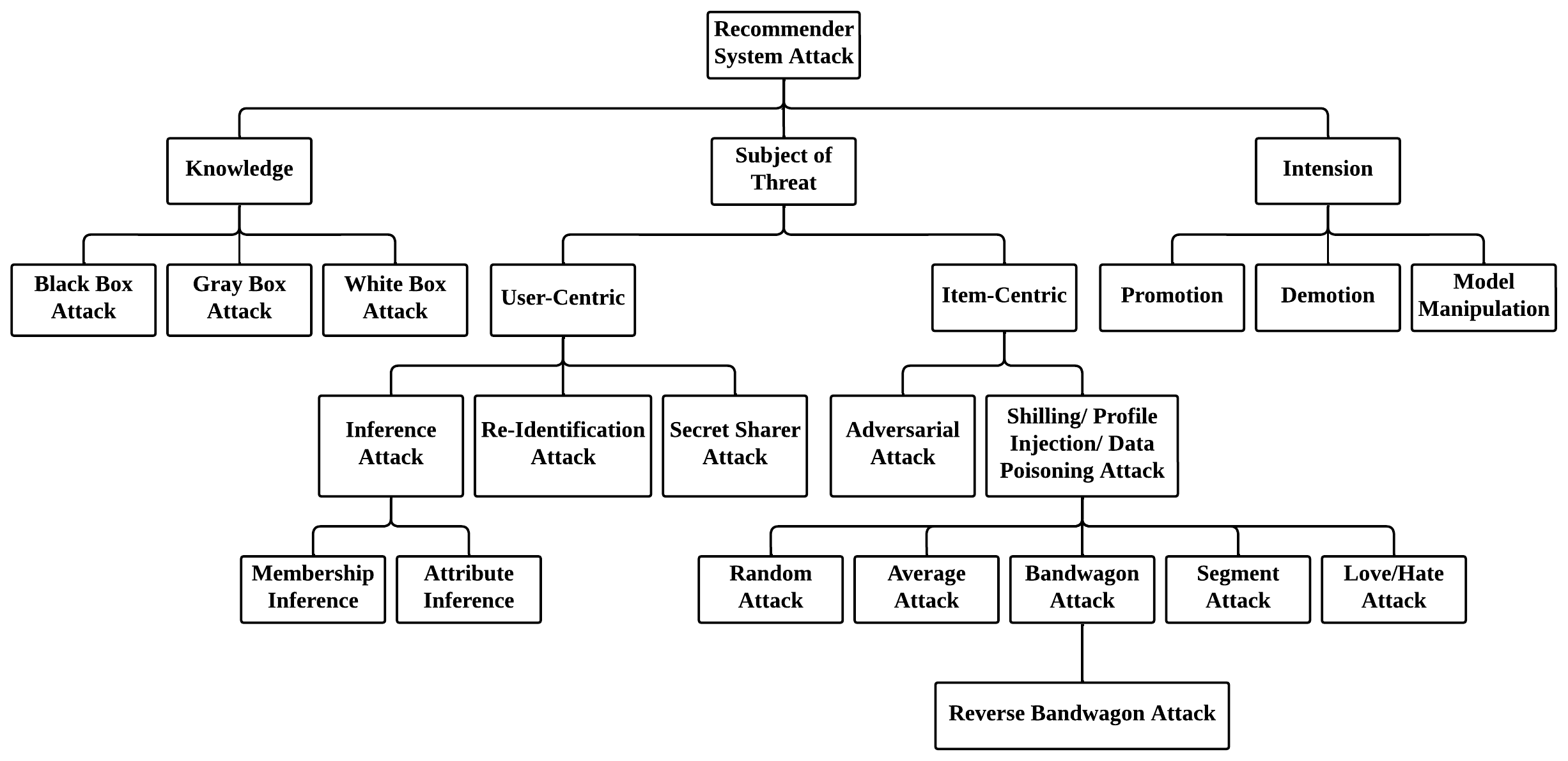}
    \caption{Recommender System Attack taxonomy}
    \label{fig:AttackClassification}
\end{figure}

Attacks can be classified based on the adversary's knowledge level into black, grey, or white box attacks \cite{fan2023adversarial, carlini2019secret, chen2022knowledge, wu2021fight, lin2022shilling, wang2022gray}. In a black box attack, the adversary can only access the recommendation output by providing inputs, with or without some auxiliary data. They cannot access the internal structure or parameters of the system \cite{fan2023adversarial, carlini2019secret, chen2022knowledge, lin2022shilling}. In contrast, a white box attack occurs when the adversary thoroughly knows the system’s internal architecture, algorithms, and parameters \cite{wu2021fight}. This detailed insight allows them to craft highly effective strategies to manipulate recommendations, degrade the system’s performance, or extract sensitive information. A grey box attack falls in between, where the adversary has partial knowledge of the system, such as the type of general algorithm used, a partial set of system parameters, or access to a subset of the training data, but not the full implementation details \cite{wang2022gray}. However, the recommender system's security and privacy features make it difficult for an adversary to have the deep knowledge required for white box or grey box attacks, making them unrealistic for real-world applications \cite{fan2023adversarial}. Conversely, a black box attack is more realistic and accessible to execute in real-life scenarios. However, it remains challenging due to the need for knowledge about the RS algorithm or architecture, though possible \cite{fan2023adversarial, carlini2019secret, chen2022knowledge, lin2022shilling, zhang2021membership, zhu2023membership, wang2022debiasing, base01_2021Gender}. Furthermore, RS attacks can be categorized as user- or item-oriented based on the nature of the threat. Each category is briefly explained in the following sections, and Figure \ref{fig:AttackClassification} shows the attack taxonomy of the recommender system.

\subsection{User-Oriented Attacks}

In this category of attacks, the adversary aims to extract users' identities, existence, or sensitive attributes by exploiting user-item interaction data to compromise users' privacy. The adversary also employs auxiliary information, which can be domain-specific or amalgamated with different domain-specific data through transfer learning or cross-domain knowledge. This category includes re-identification attacks, inference attacks, and secret sharer attacks. Re-identification attacks seek to ascertain users' identities, while inference attacks aim to determine users' existence or sensitive data. Conversely, secret sharer attacks intend to uncover and exploit private details from the system.

\subsubsection{\textbf{Re-identification Attack (RA)}}

A re-identification attack (RA) in the RS is a type of privacy exploitation where supposedly anonymized data is used to uncover the identity of individual users. RA can happen when an attacker is able to connect anonymous data with other information, thereby identifying specific users within the dataset \cite{ge2021privitem2vec}. Such attacks aim to compromise user privacy and access sensitive information.

\subsubsection{\textbf{Inference Attack (IA)}}

An inference attack in a recommender system is a malicious activity where an adversary uses the system’s outputs, such as recommendations or predictions, to infer sensitive information about the users or the underlying data. This can involve identifying whether a specific user is part of the dataset, known as a membership inference attack (MIA) \cite{zhang2021membership, zhu2023membership, wang2022debiasing}, or revealing private user attributes and preferences, termed an attribute inference attack (AIA) \cite{gong2018attribute}. Users’ private attributes typically include age, gender, political opinions, income, etc., which individuals prefer to keep confidential \cite{beigi2020privacy, base01_2021Gender}.  

Membership inference attacks (MIA) are well-known in machine learning (ML) but can also be exploited in RS to determine a user's presence. Zhang et al. \cite{zhang2021membership} first introduced the MIA attack in RS. In RS, MIA is typically conducted at the user level rather than the data sample level, distinguishing it from traditional machine learning MIA. Moreover, an adversary can use the recommendation list instead of the prediction score to attack RS. To perform MIA, an adversary needs a target and shadow recommender system, an attack model, lists of members, feature vectors, and a personalized recommendation algorithm \cite{zhang2021membership, zhu2023membership, wang2022debiasing}. The target RS is the system where the attack is executed and trained on target datasets. The shadow RS is trained on the shadow dataset and used to verify the membership existence of the target RS, generating training data for the attack model. Members are the users whose data will be used in RS for the attack. Feature vectors describe item attributes or user preferences. The attack model’s role is to determine if the target user is a member based on the dataset from the shadow RS. The adversary may also have a surrogate model to replace the target RS in the inference process, producing rich feature vectors from unidentified users who are not recognized as \textit{member} or \textit{not-a-member} in MIA \cite{zhu2023membership}. 

Conversely, attribute inference attacks (AIA) aim to extract more information than the system intends to reveal, compromising user privacy and data security. The adversary examines the RS data or the recommendation list to execute the attack. Based on user-item interaction data, the adversary exploits the relationship between user preferences, behaviours, recommendations, and data biases to infer hidden attributes like demographic information, interests, or other personal details \cite{base01_2021Gender}. These attacks exploit the system’s dependency on user data to make accurate recommendations, inadvertently exposing user information. Due to users’ lack of awareness of their privacy and the connection between public and private data, AIA can be executed in the system \cite{salamatian2015managing, jia2017attriinfer}. AIA can even successfully infer private attributes of privacy-conscious users, where some users apply privacy preservation measures while others do not within the same system \cite{jia2017attriinfer}.  

\subsubsection{\textbf{Secret Sharer Attack}} 

A Secret Sharer Attack in the RS involves an attacker subtly injecting or manipulating data from out-of-distribution to reveal sensitive information about users or items \cite{carlini2019secret}. The attacker leverages the system to uncover private details by carefully inserting crafted input data and analyzing the system’s responses from the injected data alongside similarly crafted, non-injected data. This attack poses a threat to ML-based RS, as it highlights the extent of unintended rote learning by the RS model of its input data \cite{ning2022eana}. The secret sharer attack has been employed to assess the privacy vulnerabilities of models in the absence of solid privacy assurance.

\subsection{Item-Oriented Attacks}

In this category, the adversary seeks to manipulate the recommendation algorithm to render it either ineffective or biased according to their intentions by introducing fake users into the system. The item-oriented attacks can be further divided into shilling attacks and adversarial attacks, which are listed in detail in the following: 

\subsubsection{\textbf{Shilling Attack (Profile Injection Attack/Data Poisoning Attack)}}

The shilling attack (SA) is a highly profitable attack tactic in the recommender system. It involves a malicious strategy where attackers manipulate the system’s recommendations by injecting fake user profiles targeting specific items \cite{guo2023targeted}. This type of attack is also known as \textbf{Profile Injection Attack} due to the creation of fabricated user profiles in the RS \cite{huang2023single, si2020shilling, lin2022shilling}. Additionally, since this attack poisons the RS's training data to produce biased recommendations according to the attacker’s goals, it is also referred to as a \textbf{Data Poisoning Attack} \cite{huang2023single, zhang2022pipattack, guo2023targeted, lin2022shilling, fang2020influence}. 

The fake profiles used in this attack, known as shill profiles, include feedback data such as ratings, reviews, likes, or dislikes designed to influence the recommendation of specific items artificially. The main objective of a shilling attack is to produce skewed recommendations to benefit the attacker, often for economic gain, competitive advantage, or to manipulate public opinion. The nature of the attack depends on whether the goal is to promote or demote items in the recommendations. An item promotion attack is referred to as a \textbf{push attack}, while an item demotion attack is known as a \textbf{nuke attack} \cite{huang2023single, si2020shilling, zhang2020detecting}. In a push attack, the target item's rating is always set to the highest value on the scale, whereas, in a nuke attack, the rating is set to the lowest value. Another variant of the shilling attack involves random vandalism, where the goal is to undermine the recommendation algorithm’s reliability, making the system less useful to users \cite{si2020shilling, zhang2021data}. Typically, a shill profile includes four categories of items: selected items (aligned with the target item), filler items (chosen randomly to obscure the profile), unrated items, and target items (the focus of the attack). The effectiveness of this attack is closely tied to how well the profile is crafted. Based on the profile creation strategy \cite{si2020shilling, himeur2022latest, rani2023detection}, the shilling attack is further categorized as follows:

\begin{enumerate}
    \item \textbf{Random Attack.} The shill profiles are crafted by having maximum or minimum rating values for the target items based on the attack's intent (push/nuke), where the selected items' rating values remain empty. It is also referred to as RandomBot attack \cite{zhang2021reverse}.

    \item \textbf{Average Attack.} The profiles are created by inserting the average rating value of each item in the filler set. This attack is called an AverageBot attack, too. \cite{zhang2021reverse}.

    \item \textbf{Bandwagon Attack.} In this attack, profiles are constructed with the highest rating values assigned to popular items in the system, aiming to promote these items. Conversely, a \textbf{Reverse Bandwagon Attack} is the inverse of a Bandwagon Attack; it involves creating attack profiles by selecting the least popular items and assigning them the highest rating values, with the intention of demoting these items \cite{cai2019detecting}. 

    \item \textbf{Segment Attack.} This attack requires extensive knowledge of the system users' preferences to generate fraudulent user profiles for the purpose of the attack. Initially, the attacker identifies a subset of users who are likely to seek specific items or services from the system. The attacker then creates tailored profiles that align with the preferences of this targeted user group and assigns high ratings to the item in question to promote it. This type of attack is also known as a \textbf{Target Shilling Attack} \cite{guo2023targeted}. The attack is called an \textbf{Invisible Shilling Attack} when the forged profiles and ratings are designed to seamlessly integrate with genuine user data. This method aims to manipulate the system’s recommendations while evading detection by standard anomaly detection techniques \cite{huang2023single}.

    \item \textbf{Love/Hate Attack.} It is a sub-category of a nuke attack with zero additional knowledge to demote items. In this attack, the filler items set is assigned the highest rating value, whereas the target item's rating value is set as the lowest value in the rating scale. Another intention of this attack is to degrade the RS performance. 
    
\end{enumerate}

Nevertheless, the attack can be executed by injecting a single fake user profile \cite{huang2023single}, even with the perturb and incomplete data, by applying the statistical techniques \cite{zhang2021data}. Typically, common strategies to defend against such attacks in the RS include anomaly detection, clustering, and differential privacy \cite{zhang2020detecting, zhang2020graph, rani2023detection, wadhwa2020data}.  

\subsubsection{\textbf{Adversarial Attack}}

An adversarial attack in the RS involves creating malicious feedback data to manipulate or deceive the recommendation model. These attacks exploit system algorithm vulnerabilities, resulting in inaccurate or biased recommendations. The main objectives of adversarial attacks vary, including degrading system performance or extracting sensitive information by injecting small, indistinguishable perturbations into the input data using adversarial learning techniques \cite{wu2021fight}. Deep neural network (DNN) model-based RS, such as recurrent neural networks (RNNs) and graph neural networks (GNNs), are particularly susceptible to these attacks since minor, unnoticeable changes in input data can significantly alter the output results \cite{chen2022knowledge, fan2023adversarial, anelli2021study}. To execute the attack, an adversary must create unnoticeable fake user profiles and perturb input data, known as "adversarial examples," to exploit the recommendations \cite{christakopoulou2019adversarial}. To achieve this, adversaries typically employ a min-max game strategy to facilitate adversarial learning through the use of generative adversarial networks (GANs), knowledge graph networks, or reinforcement learning techniques  \cite{christakopoulou2019adversarial, chen2022knowledge, fan2023adversarial}. The adversary attack can be termed \textbf{data poisoning attack} because the adversary injects perturbed data into the system to carry out the attack based on adversarial learning knowledge \cite{wu2021fight}. Considering the risk of adversarial attacks, adversarial learning can be utilized to detect privacy vulnerabilities within the system, helping to mitigate privacy-related threats \cite{beigi2020privacy, ganhor2022unlearning}.

However, every attack follows a specific threat model, which usually defines the attacker's knowledge, goals and impact of the attack. Table \ref{tab:threatmodel} states the general threat model of each attack by considering the black-box attack knowledge. For the thread model, we follow the structure mentioned in \cite{base01_2021Gender}.   

\begin{table}[h]
\fontsize{7.5pt}{7.5pt}\selectfont
\renewcommand{\arraystretch}{0.9}
\centering
\caption{General Threat Model of Each Attack in Recommender System Based on Black Box Knowledge}
\begin{tabular}{>{\hspace{0pt}}m{0.105\linewidth}>{\hspace{0pt}}m{0.213\linewidth}>{\hspace{0pt}}m{0.165\linewidth}>{\hspace{0pt}}m{0.202\linewidth}>{\hspace{0pt}}m{0.212\linewidth}}
\hline \\
\multicolumn{1}{>{\centering\hspace{0pt}}m{0.105\linewidth}}{\textbf{Attack}} & \multicolumn{1}{>{\centering\hspace{0pt}}m{0.213\linewidth}}{\textbf{Resources/Knowledge}} & \multicolumn{1}{>{\centering\hspace{0pt}}m{0.165\linewidth}}{\textbf{Objective}} & \multicolumn{1}{>{\centering\hspace{0pt}}m{0.202\linewidth}}{\textbf{Opportunity}} & \multicolumn{1}{>{\centering\arraybackslash\hspace{0pt}}m{0.212\linewidth}}{\textbf{Countermeasures}} \\ \\ 
\hline
\\
Re-identification & 
The adversary has access 
to a significant amount of auxiliary data, including user interactions, preferences, and historical recommendations. They may also possess computational resources to analyze and cross-reference this data. & 
The adversary aims to re-identify users by linking anonymized recommendations or interactions to specific individuals, compromising user privacy. & 
Exploit vulnerabilities in data anonymization techniques and patterns in user behaviour that can be matched with auxiliary information. The attacker capitalizes on similarities between anonymized data and known user profiles. & 
Implement robust differential privacy with anonymization techniques, ensure data is sufficiently anonymized, and apply noise to user interactions \cite{shakil2021towards}. \\ 

Membership Inference (MIA) & 
The adversary has access to the recommender system's outputs and can observe the recommendations provided to users. & 
It aims to determine whether a specific user's data was included in the RS training dataset. & 
Leverages the patterns and anomalies in the system's outputs, exploiting differences in recommendations based on whether a user's data is part of the training dataset. &  
Implement differential privacy \& randomization techniques to ensure that the inclusion or exclusion of any single user's data has minimal impact on the system's outputs, thereby reducing the effectiveness of MIA \cite{zhang2021membership}. \\ \\

Attribute Inference (AIA) & 
The adversary has access to the recommendation outputs and possibly some auxiliary information about the users or items. The adversary also might have access to the user-item interaction data. &  
Motives to infer sensitive user attributes, such as demographics, political views, or preferences, based on the patterns observed in the recommendations. & 
The recommender system's outputs reflect user-specific data and patterns, which the adversary can analyze to extract private information. & 
Implement different randomization techniques to obscure the influence of individual user data in the recommendations, reducing the risk of attribute inference \cite{base01_2021Gender, salamatian2015managing, lin2022privacy, weinsberg2012blurme, feng2015can}.\\ \\

Shilling (or Data Poisoning or Profile Injection) & 
The adversary possesses the capability to create fake user profiles and generate biased data to manipulate the recommendation list. & 
Promote (push) or demote (nuke) specific items in a false manner within the recommendation list, manipulating user preferences and behaviour. & 
Exploits the system's reliance on user-generated data, taking advantage of the system's inability to distinguish between genuine and crafted profiles. & 
Implementing robust detection algorithms, such as anomaly detection and profile clustering, along with differential privacy, can block the adversary from conducting the attack \cite{zhang2020detecting, zhang2020graph, rani2023detection, wadhwa2020data}.
\\ \\

Adversarial & 
The adversary has a deep understanding of the RS's algorithms and access to its input data. They can generate and inject specially crafted malicious inputs to achieve their goals. & 
Degrade the system's performance or extract sensitive information by misleading the system into making incorrect predictions or recommendations. & 
Leverages the inherent vulnerabilities in the recommendation algorithms. & 
Employing robust machine learning techniques with privacy preservation, such as classification, adversarial training, regularization, \& GAN \cite{cao2020adversarial, beigi2020privacy, ganhor2022unlearning}.
\\ 

Secret Sharer & 
The adversary has a deep understanding of the RS’s algorithms and access to its input or output data. They may also have the capability to inject or manipulate data within the system. & 
The goal is to exploit the privacy of the PPRS & 
Leverages the system’s predictive capabilities, which can inadvertently expose private details through patterns in the recommendations or the inclusion of specific data & 
Implementing strong privacy-preserving techniques such as differential privacy to mitigate the impact of unintended memorization of the model \cite{ning2022eana}.\\
\hline
\label{tab:threatmodel}
\end{tabular}
\end{table}

\section{Privacy of Recommender System}
\label{sec:privacy}

In the recommender system context, accessing more of a user's previous data allows the system to generate better recommendations. However, this also increases the risk of privacy breaches. Generally, there is an inverse relationship between the accuracy of the RS and the users' privacy. Much research focuses on balancing system accuracy and users' privacy. In RS, privacy means keeping users' activities confidential, such as their feedback data. This feedback data includes actual ratings and the list of rated items which should stay private. Protecting this data is known as data leakage protection \cite{wang2021cryptorec}. Additionally, users' feedback data is closely linked to their private attributes like gender and age, which also requires protection from inference attacks \cite{wang2021cryptorec}.

\begin{figure}[ht]
\includegraphics[width=0.95\linewidth]{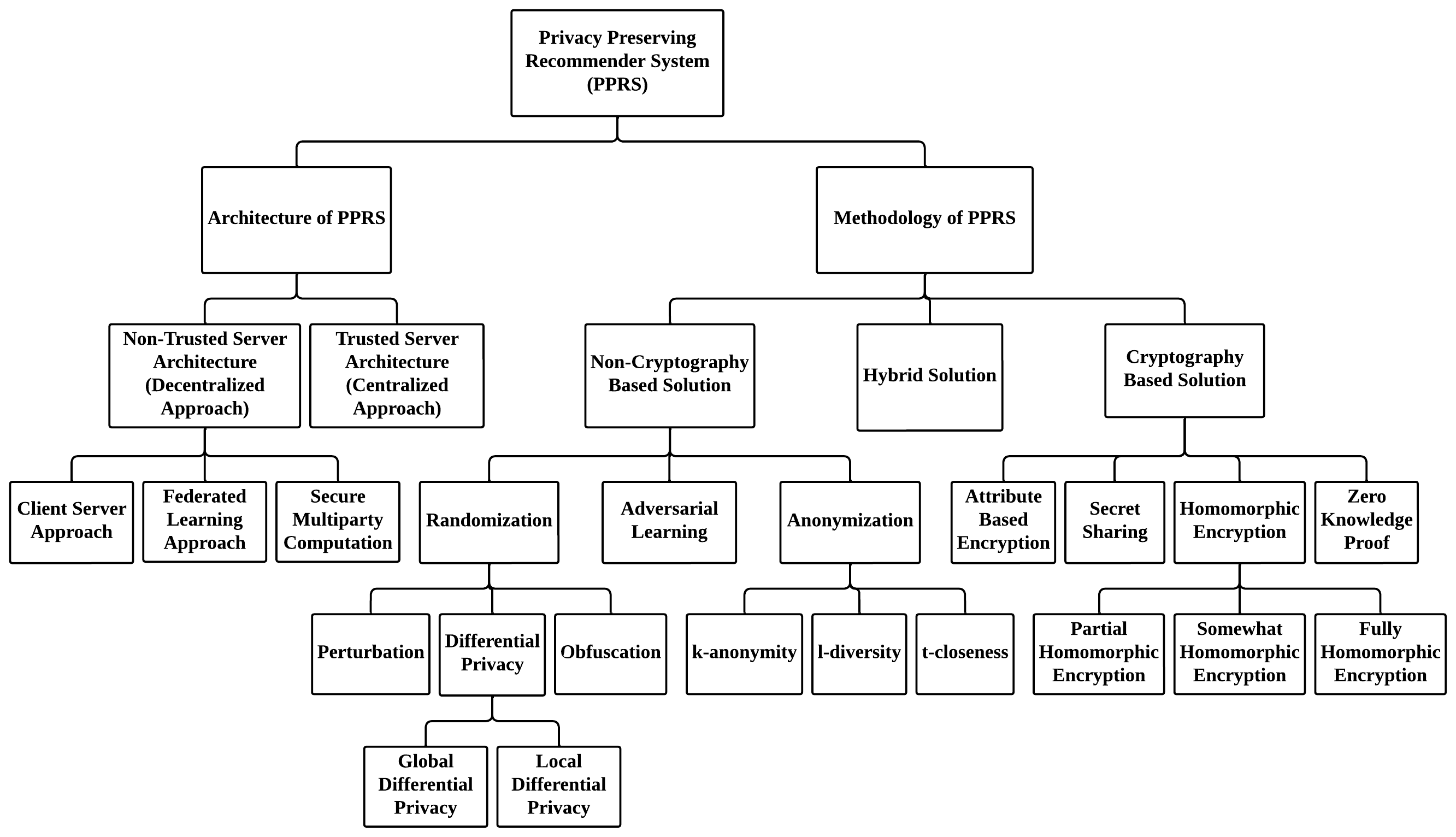}
\caption{A Taxonomy of Privacy-Preserving Recommender System (PPRS)}
\label{fig:pprs}
\end{figure}

However, the RS's privacy preservation techniques can broadly be categorized into cryptography, non-cryptography, and hybrid approaches. Implementing these techniques also depends on the system's architecture, which is further classified into two categories based on server characteristics. Figure \ref{fig:pprs} depicts the elaborate view of the privacy preservation recommender system (\textit{PPRS}). 

The RS collects users' feedback data and uses it in the algorithm to provide recommendations. The system then stores this information. To be effective, the RS must gain users' trust so they provide accurate data. Without user trust, the RS may not get genuine feedback, which can affect its business value \cite{palomares2021reciprocal}. Therefore, the system must gain users' reliability by ensuring data privacy and the privacy preservation techniques depend on the system architecture. The consequent sections \ref{PPRS:Architecture} \& \ref{PPRS:Method} explain the structure and applied methodology of privacy-preserving RS to ensure users' privacy.

\subsection{Architecture of Privacy-Preserving Recommender System}
\label{PPRS:Architecture}

A privacy-preserving recommender system (PPRS) can be implemented using a centralized or decentralized architecture based on the system's reliability. In a centralized setup, the system is a trusted server responsible for maintaining data privacy \cite{liu2024matrix, mullner2023reuseknn, yang2023fairness, chen2021differentially, beigi2020privacy}. The \textbf{centralized recommender system} gathers and stores users' rating data and maintains user profiles to generate recommendations  \cite{wang2018toward}. The server has two primary responsibilities: providing precise \& accurate recommendations and protecting user data privacy by ensuring that no personal information can be inferred from the recommendations. To achieve privacy protection, centralized RS typically uses non-cryptographic techniques due to their ease of application and low computational complexity \cite{liu2024matrix, mullner2023reuseknn, yang2023fairness, ogunseyi2023systematic}. However, this architecture is suitable for users who are not aware of data privacy concerns. Usually, privacy perceptions vary among users; some may consider specific data private, while others do not. Consequently, users' private data can be inferred using the publicly shared data of others who do not consider it private \cite{jia2017attriinfer}. In this context, the centralized architecture effectively preserves and ensures users' data privacy. Furthermore, centralized RS offers better accuracy in recommendations because the system solely collects and processes users' and items' data. 

Conversely, a decentralized architecture is appropriate for preserving users' data privacy when the system is not entirely trustworthy or is semi-trustworthy. In such systems, privacy preservation is the user's responsibility. A semi-trustworthy server is honest but curious about gaining private information without user consent. Additionally, a server can undergo a "mission creep," starting honestly but later changing its business strategy to infer or collect users' private data. Alternatively, it may be transferred to another company that attempts to do so \cite{base01_2021Gender}. In a decentralized architecture, user profiles and rating data are stored on the user end, and users are fully responsible for ensuring their privacy. This architecture, also known as a user-centralized architecture \cite{zhang2019caching}, assures better privacy than centralized RS, as users actively manage their data during the recommendation process \cite{ogunseyi2023systematic}. 

Decentralized RS can be further categorized into \textbf{Distributed Client-Server (DCS), Secure Multiparty Computation (SMC)}, or \textbf{Federated Recommender System (FedRec)} based on where the recommendation model executes. In DCS-based RS, users and non-trusted systems act as clients and servers, respectively, where clients request services from the server. Clients first modify their data in a privacy-protection mode and then send the protected data to the server for processing and service delivery \cite{ben2021privacy, neera2021private, hu2021rap, bao2021privacy, zheng2023decentralized}. In contrast, SMC-based RS allows multiple parties, such as a group of users, to collaboratively compute a function over their inputs while keeping those inputs private. The primary goal is to ensure that no user gains information about others' inputs beyond what can be inferred from the output. SMC relies on cryptographic techniques such as secret sharing \cite{chen2021secrec, ben2021privacy} and homomorphic encryption \cite{zhang2021privacy, zhang2022homomorphic} to ensure input confidentiality during computation. Computation is performed interactively, with parties exchanging encrypted data to compute the final result jointly.

FedRec, primarily an ML-based RS, follows federated learning (FL) principles, where users manage their data and use a pre-trained global recommendation model to produce recommendations on their devices. Instead of sending actual data, users send trained model updates to a central server, where updates can be gradients or weights that are smaller than the entire dataset. The central server aggregates updates from all participating devices, improves the global model, and redistributes it to all user devices. The process of local training, sending updates, and global aggregation is repeated periodically, allowing the model to continuously learn and adapt to new data while maintaining user privacy \cite{yang2020federated}. According to the component sharing, the FedRec can be either a horizontal federated recommender system (HFedRec), a vertical federated recommender system (VFedRec), or a transfer federated recommender system (TFedRec) \cite{yang2020federated, wang2024horizontal}. HFedRec trains the model by integrating users' historical data on shared items from different users without compromising individual privacy. VFedRec trains a model by utilizing auxiliary information from different RSs in a privacy-preserving manner. TFedRec trains a model by transferring knowledge from source to target domain parties without revealing user privacy \cite{yang2020federated, yang2019federated, wang2024horizontal}. 

The privacy requirements of decentralized RS are similar to those of centralized architecture, but users are more actively involved in ensuring their data security. Although decentralized RS inherently provides a privacy protocol, as users preserve their entire historical data, and even FedRec prioritizes user privacy over DCS, it remains vulnerable to threats such as inference attacks and differential attacks \cite{chai2022efficient, hu2024user, yin2024poisoning, zhang2023comprehensive, yuan2023manipulating, zhang2022pipattack}. To address these vulnerabilities, various privacy preservation methodologies, such as randomization, differential privacy, secure multiparty computation, and homomorphic encryption, can be applied to build decentralized PPRS \cite{ben2021privacy, neera2021private, hu2021rap, bao2021privacy, zheng2023decentralized, chai2022efficient, hu2024user, zhang2023comprehensive}. Also, to maintain strict data privacy, hybrid RS architectures can be implemented by combining multiple approaches, such as incorporating secure multiparty computation into a FedRec structure \cite{zhang2022homomorphic}.  Table \ref{tab:ArchiPPRS} lists the advantages and disadvantages of centralized and decentralized PPRS architectures.

\begin{table}[h]
\fontsize{7.5pt}{7.5pt}\selectfont
\renewcommand{\arraystretch}{1.1}
\centering
\caption{List of Pros \& Cons of Different Architecture of Privacy-Preserving Recommender System}
\label{tab:ArchiPPRS}
\begin{tabular}{>{\centering\hspace{0pt}}m{0.125\linewidth}>{\hspace{0pt}}m{0.325\linewidth}>{\hspace{0pt}}m{0.470\linewidth}}

\hline 
\textbf{Architecture} & \multicolumn{1}{>{\centering\hspace{0pt}}m{0.325\linewidth}}{\textbf{Pros}} & \multicolumn{1}{>{\centering\arraybackslash\hspace{0pt}}m{0.470\linewidth}}{\textbf{Cons}} \\ 
\hline  

 & & \\
Centralized & 
\begin{tabular} {@{\labelitemi\hspace{\dimexpr\labelsep+0.5\tabcolsep}}l@{}}Suitable where the privacy definition\end{tabular}\par{}varies user to user.\par
\begin{tabular}{@{\labelitemi\hspace{\dimexpr\labelsep+0.5\tabcolsep}}l@{}}More applicable where users are not ~\end{tabular}\par{}aware of the linkage between their public and private data.\par
\begin{tabular}{@{\labelitemi\hspace{\dimexpr\labelsep+0.5\tabcolsep}}l@{}}Users’ active participation is not req$-$  ~\end{tabular}\par{}uired to acquire the recommendation.\par
\begin{tabular}{@{\labelitemi\hspace{\dimexpr\labelsep+0.5\tabcolsep}}l@{}}Cost-effective as it does not require any  ~\end{tabular}\par{}communication cost between server and client.\par
\begin{tabular}{@{\labelitemi\hspace{\dimexpr\labelsep+0.5\tabcolsep}}l@{}}Provide more accuracy in the recom$-$  ~\end{tabular}\par{} mendation as the system solely stores all the data and does all the required computations at the system end.

& \begin{tabular}{@{\labelitemi\hspace{\dimexpr\labelsep+0.5\tabcolsep}}l@{}}The system itself can be vulnerable or mission creep. \\ Might suffer man-of-the-middle attack. \\Also, a user might be an adversary who can conduct active. \end{tabular}\par{} inference and background knowledge attacks from the recommendation he receives from the system. 

\\ \hline 
 & & \\
Decentralized & 
\begin{tabular}{@{\labelitemi\hspace{\dimexpr\labelsep+0.5\tabcolsep}}l@{}}Guarantees better privacy and\end{tabular}\par{}personalization than centralized one.\par
\begin{tabular}{@{\labelitemi\hspace{\dimexpr\labelsep+0.5\tabcolsep}}l@{}}Users’ participation also increases users' \end{tabular}\par{}awareness about their privacy.\par
\begin{tabular}{@{\labelitemi\hspace{\dimexpr\labelsep+0.5\tabcolsep}}l@{}}Reduce the risk of data breaches and \end{tabular}\par{} unauthorized access.\par
\begin{tabular}{@{\labelitemi\hspace{\dimexpr\labelsep+0.5\tabcolsep}}l@{}}For FedRec, it is easy to imply privacy \end{tabular}\par{} regulations like GDPR and CCPA, which impose strict rules on data storage and processing.\par

& \begin{tabular}{@{\labelitemi\hspace{\dimexpr\labelsep+0.5\tabcolsep}}l@{}}Implementing a decentralized approach is more complex\end{tabular}\par{} than traditional centralized ones as it causes much communication overhead between the client and server with the computational overhead at the client side to preserve their data privacy which overall increases the cost in comparison with the centralized approach.\par
\begin{tabular}{@{\labelitemi\hspace{\dimexpr\labelsep+0.5\tabcolsep}}l@{}}User-end overhead might harm users satisfaction level.\end{tabular}
\begin{tabular}{@{\labelitemi\hspace{\dimexpr\labelsep+0.5\tabcolsep}}l@{}}Variability in device capabilities and data distributions can\end{tabular}\par{} complicate model training and consistency. \par
\begin{tabular}{@{\labelitemi\hspace{\dimexpr\labelsep+0.5\tabcolsep}}l@{}}Different users might have different privacy concerns that \end{tabular}\par imply a different level of privacy preservation, and it might harm the system’s accuracy in providing actual recommendations. \par
\begin{tabular}{@{\labelitemi\hspace{\dimexpr\labelsep+0.5\tabcolsep}}l@{}}The FedRec ensure the protection of users’ rated items' \end{tabular}\par{} value privacy by allowing them to function the model locally, but it could not assure the existence privacy at the time of sharing gradients \cite{zheng2023federated}. \par
\\ \hline
\end{tabular}
\end{table}

\subsection{Methodology of Privacy-Preserving Recommender System}
\label{PPRS:Method}

Based on privacy preservation techniques, the PPRS can be categorized as cryptography, non-cryptography \& hybrid techniques, which are described in detail in the following sections. 

\subsubsection{\textbf{Cryptography-based Methodology}}

Cryptography-based methodology in RS involves using encryption and related techniques to ensure the security and privacy of user data in the recommendation process. This methodology assures that sensitive user data remains protected during data collection, processing, and recommendation generation. It includes encrypting user preferences and leveraging cryptographic protocols to protect sensitive information. Incorporating cryptography into RS can enhance data privacy and security, ultimately leading to more trustworthy and reliable recommendations. A wide range of cryptography techniques exist, but based on the nature of RS, the following cryptography-based methodologies can be applied to ensure security and privacy. 

\begin{itemize}

    \item \textbf{Homomorphic Encryption (HE)}, the most widely used cryptography-based approach in RS, offers diverse computational capabilities of the ciphertext without decryption \cite{zheng2023federated}. This method can be implemented as a client-server structure or integrated into the secure multiparty computation (SMC) approach \cite{badsha2017privacy, kim2018efficient, zhang2021privacy}. In the HE methodology, users encrypt their rating data before sending it to the server (system) to receive the recommendation services. The server conducts computations on it without decrypting the actual value of the ciphertext to generate the recommendations \cite{li2020privacy, acar2018survey}.
    These computations, usually addition or multiplication, are equivalent to performing the same operations on the plaintexts by the client end, as demonstrated in the mathematical form by Equation \ref{eqn:HE} \cite{li2020privacy}.


    \begin{equation}
    \label{eqn:HE}
        \begin{matrix}
         E(m_1) \oplus E(m_2) = E(m_1 + m_2) \approx     D(E(m_1) \oplus E(m_2)) = m_1 + m_2    \\
         E(m_1) \otimes  E(m_2) = E(m_1 . m_2) \approx     D(E(m_1) \otimes  E(m_2)) = m_1.m_2 \\
         m_1, m_2 \epsilon M
        \end{matrix}
    \end{equation}

    Here, $E(.)$ \& $D(.)$ denote corresponding encryption \& decryption algorithm. $m_1$ \& $m_2$ indicate users' ratings and $M$ defines the rating set. However, $\oplus$ and $\otimes$ determine the addition \& multiplication operations on ciphertexts, respectively.  

    
    Based on the extent of computational operations that can be performed, the HE can be classified as either partial homomorphic encryption (PHE), somewhat homomorphic encryption (SWHE), or fully homomorphic encryption (FHE). The partial homomorphic encryption method supports an unlimited number of a single type of operation (addition or multiplication, but not both) on the ciphertext. Cryptosystems like Paillier and ElGamal are commonly utilized in PPRS to implement PHE for privacy preservation \cite{ge2021privitem2vec, zhang2022homomorphic, niu2023federated, badsha2018privacy}. Meanwhile, the somewhat homomorphic encryption allows for unlimited addition operation but a limited number of multiplications on the ciphertext \cite{park2020shecs, badsha2017privacy, zhang2021privacy, badsha2018privacy}. Boneh Goh Nissim (BGN) cryptosystem is used to implement the SWHE technique \cite{badsha2017privacy, zhang2021privacy, badsha2018privacy}. Fully homomorphic encryption allows arbitrary addition \& multiplicative computations on ciphertexts, and it is more suitable in PPRS \cite{jumonji2021privacy, zhou2023lightweight, kim2018efficient}. Although the high computational cost of FHE can sometimes render it impractical for real-world applications, ongoing research aims to mitigate these costs by using a packing mechanism \cite{jumonji2021privacy} or proposing new data structures, such as vectors, to encrypt data and perform computations on ciphertext vectors \cite{kim2018efficient}. 
    \\
    \item \textbf{Secret Sharing (SS)} methodology can be applied instead of HE to reduce communication overhead \cite{zhang2024hn3s, 2022FakeMarksSS, zheng2023federated, chen2021secrec}. In the SS technique, each participant divides a rating value into $n_i$ random numbers. Afterwards, every participant keeps one copy of a random number and shares $N = n_i-1$ copies with other $J$ participants. Then, a participant adds all the received copies of the share with the single copy of their share as $\theta_i = (n_i - N) + \sum_{j=1}^{J} Share(n_j)$ and sends the addition value to the server. The server collects and accumulates all the gradients from the participants. Usually, the accumulated gradient is equivalent to the gradient without applying SS. Later, the server transfers the updated gradient to the participants, where they update their gradient by adding as $\theta_i = \theta_i + \theta_{server}$. The SS methodology can protect existence privacy by ensuring that the server only collects a collection of gradients from a set of users, preventing it from inferring the actual preferences of any individual user. This approach is also well-suited for FedRec, where minimizing communication overhead is another key concern alongside privacy \cite{zhang2024hn3s, 2022FakeMarksSS, zheng2023federated}. 
    \\

    \item \textbf{Attribute-Based Encryption (ABE)} technique allows for fine-grained access control of encrypted data, which is suitable for sharing data with the non-trusted server for getting services \cite{huan2022privacy, yin2019privacy, wu2023privacy}. In ABE, a group of users possesses a set of attributes that serve as their private keys. These attributes are linked to specific functions that determine each user's ability to decrypt encrypted texts. Based on access control policy, ABE can be classified as key-police ABE (\textbf{KP-ABE}), ciphertext-police ABE (\textbf{CP-ABE}), anti-quantum ABE (\textbf{AQ-ABE}) and generic ABE (\textbf{G-ABE}) \cite{zhang2020attribute}. Among them, CP-ABE is extensively used in PPRS with other encryption techniques such as search encryption or secret sharing due to the continuous information exchange nature of RS to acquire necessary services \cite{huan2022privacy, yin2019privacy, wu2023privacy}. \\

    \item \textbf{Zero-Knowledge Proof (ZKP)} is another cryptographic protocol involving an authentication process between two parties: the prover and the verifier. In this protocol, the prover confirms a statement's validity to the verifier without revealing any details about the statement itself. This process ensures the truthfulness of the statement without disclosing specific information. In other words, the prover can demonstrate the accuracy of a statement without revealing the actual content. ZKP achieves data privacy through a series of interactions between the prover and the verifier, where the prover provides evidence of knowledge without exposing the knowledge itself \cite{yin2023zero}. This concept requires the user to prove that their actions adhere to a specified protocol, ensuring honest behaviour while maintaining privacy \cite{ogunseyi2023systematic}. 
    
 \end{itemize}

Table \ref{tab:PPRS_Crypto} demonstrates the summary of reviewed papers on the cryptography-based methodologies applied in the RS to ensure data privacy by considering a specific problem scenario. 

\begin{table}[h]
\centering
\fontsize{7.5pt}{7.5pt}\selectfont
\renewcommand{\arraystretch}{0.9}
\caption{Summary of Cryptography-based Privacy-Preserving Recommender System}

\begin{tabular}{>{\centering\hspace{0pt}}m{0.084\linewidth}>{\centering\hspace{0pt}}m{0.184\linewidth}>{\centering\hspace{0pt}}m{0.228\linewidth}>{\centering\hspace{0pt}}m{0.205\linewidth}>{\centering\arraybackslash\hspace{0pt}}m{0.180\linewidth}} 
\hline
\textbf{Reviewed Paper} & \textbf{Privacy Methodology} & \textbf{Considered Problem Scenario} & \textbf{Considered Performance Parameter} & \textbf{Experimental Domain \& Dataset Name} \\ 
\hline 

\cite{zhang2022homomorphic} & HE $\to$ PHE\par{}(Additive ElGamal cryptosystem) & Data leakage caused in FL environment by model reconstruction attack or model inversion attack\par{} & RS Accuracy, Security, Communication \&  Computational Cost & Health (HAM10000)\tablefootnote{\label{note:HAM}https://dataverse.harvard.edu/dataset.xhtml?persistentId=doi:10.7910/DVN/DBW86T} \\ 

\cite{ge2021privitem2vec} & HE $\to$ PHE\par{}(Paillier cryptosystem) & Data Leakage, Computation \& Communication Overhead & Privacy-Utility trade-off \& Efficiency\par{} & Movie (ML@100k\tablefootnote{\label{note:ML}https://grouplens.org/datasets/movielens/}) \\ 

\cite{zhang2021privacy} & HE $\to$ SWHE \par{}(BGN Cryptosystem) & AIA & Computational Cost \&\par{}Communication Overhead & Health (UCI\par{}ML Datasets\tablefootnote{\label{note:UCI}https://archive.ics.uci.edu/}) \\

\cite{badsha2017privacy} & HE $\to$ SWHE \par{} (BGN cryptosystem) & Sharing other users ciphertext to the target users to conduct multiplication operation of HE\par{} & Security, Communication \&  Computational Cost & Movie (Customized ML@)\footref{note:ML}\\

\cite{kim2018efficient} & HE $\to$ FHE \par{} (RLWE schemes) & Users Privacy \& the bottleneck of the gradient descent computation of FHE in CF\par{} & RS Accuracy \& Efficiency of FHE & Movie (ML@100k)\footref{note:ML} \\

\cite{niu2023federated} & HE $\to$ PHE \par{} (Paillier cryptosystem) & Information leakage from gradients sharing in FL environment & RS accuracy \& Efficiency (Time \& Convergence) & Movie (ML@1M)\footref{note:ML} \\

\cite{badsha2018privacy} & HE $\to$ PHE, SWHE (Paillier \& BGN cryptosystem) & Data leakage \& security risks of storing sensitive data & RS accuracy, Computation \& Communication overhead\par{} & POI (WSDREAM\tablefootnote{\label{note:QoS}https://wsdream.github.io/}) \\

\cite{jumonji2021privacy} & HE $\to$ FHE \par{} (BGV cryptosystem) & Communication Traffic \& Chosen Plaintext Attack & Security, Communication \&  Computational Cost\par{} & Movie (ML@)\footref{note:ML} \\

\cite{zhou2023lightweight} & HE $\to$ FHE \par{} (tag-based multikey cryptosystem) & Chosen Ciphertext Attack & RS accuracy, scalability \\ Communication \&  Computational Cost & Movie (ML@100k)\footref{note:ML} \\

\cite{zheng2023federated} & Secret Sharing & Mainly focused on existence privacy & Privacy-Utility trade-off \& communication overhead & Movie (ML@100K\footref{note:ML}, FilmTrust\tablefootnote{\label{note:FlimT}https://guoguibing.github.io/librec/datasets.html}) \& E-commerce (Epinion\tablefootnote{\label{note:EP}http://konect.cc/networks/epinions/}) \\

\cite{chen2021secrec} & Secret Sharing & Privacy leakage from gradient sharing \& considering offline users scenario\par{} & Computational \& Communication overhead and RS accuracy & Movie (ML@1M\footref{note:ML}) \\

\cite{2022FakeMarksSS} & Secret Sharing \& Fake Marks &  Disclosure of rating value \& rating behaviour history & Privacy-Utility trade-off \& communication overhead & Movie (ML@100K \& ML@1M)\footref{note:ML} \\

\cite{zhang2024hn3s} & Secret Sharing & Gradient attack \& non-active users in FL environment & Privacy-Utility trade-off \& communication overhead & Academic (Citeulike-a\tablefootnote{\label{note:Cite}https://github.com/js05212/citeulike-a}), Gaming (Steam\tablefootnote{\label{note:Steam}https://www.kaggle.com/datasets/tamber/steam-video-games}), Movie (ML@1M\footref{note:ML}, FilmTrust\footref{note:FlimT})\par{} \\ 



\cite{huan2022privacy} & ABE \& Search Encryption & Lack of personalization recommendations in e-leaning \& teachers privacy with secure key management & Privacy-Utility trade-off & Education (Synthetic Dataset) \\ 

\cite{yin2019privacy} & ABE \& Search Encryption & Sensitive data exposure & Efficiency-Personalized recommendation trade-off \& operational cost\par{} & Job (Synthetic Dataset)  \\

\cite{wu2023privacy} & ABE \& Secret Sharing & Sensitive information leakage while matching friends & Privacy-Utility trade-off, Computational overhead, Execution efficiency, Adaptability \& Flexibility\par{} & Social Network (Synthetic Dataset\tablefootnote{\label{note:SN}Dataset can be available upon the request of the corresponding author of the paper.}) \\ 

\cite{yin2023zero} & Zero-Knowledge Proof & Potential data leakage for acquiring personalized recommendation & Computational overhead, signature generation optimization, \& security measures & Education (Synthetic Dataset) \\ 

\hline
\end{tabular}
\label{tab:PPRS_Crypto}
\end{table}

\subsubsection{\textbf{Non-Cryptography-based Methodology}}
\label{sec: noncrp}

The non-cryptography-based methodology is aimed at protecting and preserving users' privacy by either removing sensitive personal attributes or adding noise to conceal personally identifiable information in the dataset. These methods, which do not involve any cryptography protocol, ensure that the user's data and preferences remain confidential. Importantly, they also enable the RS to provide accurate and relevant suggestions. Instead of relying on cryptography operations, these techniques focus on data transformation, aggregation, and perturbation to achieve privacy objectives. The non-cryptography-based methodology is broadly categorized as anonymization, adversarial learning, and randomization, as described in the following section. 

\begin{itemize}

    \item \textbf{Anonymization} involves concealing or eliminating personally identifiable information from data before it is shared, ensuring individuals' privacy. Common anonymization methods include k-anonymity, l-diversity, and t-closeness \cite{zhang2019caching, zhao2018illia, mehta2022improved}. \textbf{K-anonymity} is a well-known and pioneering privacy preservation technique that ensures each data entry in the dataset has at least \textbf{k-1} other entries with the same values for quasi-identifiers, which are attributes that can identify a person when combined. K-anonymity can be achieved using suppression or generalization techniques. This technique initially generates clusters of various lengths based on users' ratings and total item sets. Then, user clusters are formed so that each user's data aligns with precisely k-1 other users' data, thus safeguarding privacy \cite{zhao2018illia, zhang2019caching, wei2018improving}. However, k-anonymity cannot guarantee complete privacy and is vulnerable to homogeneity, background knowledge attacks, and linkage attacks, especially when there is insufficient diversity among sensitive values in the dataset \cite{machanavajjhala2007diversity, parameshwarappa2021anonymization, valdez2019users, saleem2021parking}. To address these shortcomings, Machanavajjhala et al. \cite{machanavajjhala2007diversity} proposed l-diversity, which ensures that each sensitive attribute of the user should have \textbf{\textit{l}} distinct values for every equivalence class, which share the same quasi-identifier values. Nevertheless, l-diversity alone cannot entirely prevent attribute disclosure \cite{li2006t, mehta2022improved}. To enhance anonymization, the \textbf{t-closeness} technique was introduced. It asserts that the distribution of a sensitive attribute in any equivalence class should closely match the distribution of that attribute in the overall dataset, with the threshold \textbf{\textit{t}} maintaining this closeness \cite{li2006t, parameshwarappa2021anonymization}. However, combining all three anonymization techniques strengthens anonymization and reinforces privacy preservation \cite{wei2018improving}.          
    \\
    
    \item \textbf{Adversarial Learning (AL)} is commonly used in ML to train the models to withstand adversarial examples. These adversarial examples are intentionally crafted to mislead the model. \cite{lowd2005adversarial}. In PPRS, the AL technique can be implemented with two components: the recommendation module and the inference attack module \cite{beigi2020privacy}. Based on the user's historical data, the recommender module generates recommendations as a ranked list of approximate preferred items for the target users. Meanwhile, the inference attack module tries to gain information about the users' private attributes from the recommendation items. Finally, the users' privacy is maintained by optimizing the min-max game between modules and reconstructing the recommendation list in a way that the adversary can not infer users' private attributes. Equation \ref{eq:AL} is the mathematical formulation of the min-max game between the modules.

    \begin{equation}
    \label{eq:AL}
        RAP = min_{\theta_R}(\boldsymbol{L}_R - \alpha \;\;  max_{\theta_P} \;(\boldsymbol{L}_P))
    \end{equation}

    Here, $\boldsymbol{L}_P$ \& $\boldsymbol{L}_R$ denote the information gain \& recommendation error of inference attack \& recommender module, respectively. Also, $\theta_R$ indicates the set of embedding matrices for users \& items. $\theta_P$ states all the required parameters of the inference attack module to conduct the attack and gain private attribute information. However, $\alpha$ controls the effect of the attack module in the learning process. Additionally, the presence of bias in the system results in privacy breaches \cite{ganhor2022unlearning}. AL can be employed to identify and understand the bias in recommendations, allowing for the reduction and control of biases to manage users' exposure to implicit private information.   
    \\

    \item \textbf{Randomization} is another non-cryptographic technique used to protect privacy in recommender systems. This approach involves altering data by introducing uncertainty or noise. Randomization can be divided into three sub-categories based on the type of data alteration: perturbation, obfuscation, and differential privacy. \\

    \begin{itemize}
        \item \textbf{Perturbation}. In this approach, users' data are altered by adding noise to the actual data in a controlled way, ensuring the RS's utility remains unaffected while maintaining user privacy. The noise is added by the actual data with random values in a controlled manner \cite{polatidis2017privacy}, such as, selecting noise from Gaussian distribution \cite{yalcin2024novel}, or by applying chaos theory for noise injection \cite{kashani2020feature} or mixing users' private ratings with the available public services' ratings based on the required privacy level of the users \cite{hu2021rap}. Perturbation ensures data privacy to a certain level but cannot provide any guarantee as it depends on noise addition in the actual data. Perturbation makes it difficult, but not impossible, for an adversary to infer the exact value. An adversary can potentially retrieve the actual data with high precision by analyzing the clustered perturbed data \cite{xu2017privacy}. Additionally, the recommendation accuracy may be compromised due to the added noise \cite{polatidis2017privacy}. However, to improve the privacy assurance of the perturbation approach, it can be implemented by integrating with other methods, such as k-anonymity \cite{kashani2020feature}, differential privacy \cite{liu2017differential}.  \\

        \item \textbf{Obfuscation}. In the obfuscation approach, the data privacy disclosure risk is mitigated by distorting the data by adding, removing, or altering existing data in a way so that the whole process blocks the adversary from conducting an inference attack and also ensures recommendations accuracy as it contains no such noisy data \cite{lin2022privacy}. The obfuscation can be achieved by solving the multi-objective optimization problem where the objectives are improving RS accuracy \& protecting users' privacy \cite{feng2015can} or applying probabilistic privacy mapping between private \& to be released public data \cite{salamatian2015managing} or removing the most indicative items of the sensitive data with the addition of some fictitious data to confuse the adversary \cite{base01_2021Gender, lin2022privacy, weinsberg2012blurme}. However, this approach can not provide a privacy guarantee and also suffers differential attack as it distorts the data with actual existing other data. \\

        \item \textbf{Differential Privacy (DP)} is a widely applied privacy preservation technique that ensures robust user privacy protection. In PPRS, DP is implemented by adding noise in a way that makes the inclusion or exclusion of a single user’s data have an indistinguishable impact on the output of the recommendation algorithm. This carefully calibrated noise can be introduced at various stages: on the input data \cite{mullner2023reuseknn, neera2021private, chai2022efficient, zhang2021graph}, the model parameters \cite{wang2023improved, liu2017differential, ning2022eana}, or the final recommendations \cite{zhang2021graph}. DP safeguards individual user privacy even when an adversary possesses significant auxiliary information about the data or the algorithm's execution process. It enables the recommendation system to produce valuable recommendations while maintaining robust privacy assurances, ensuring a fair balance between data utility and privacy protection. DP offers a rigorous mathematical definition of privacy preservation with a lower computational burden than cryptographic methods. Initially, differential privacy is denoted as $\varepsilon-$differential privacy ($\varepsilon-$DP) which can be defined as follows:
        
        \begin{mydef}
        \label{def:DP1}
        \textbf{$\varepsilon-$Differential Privacy}. 
        A randomized computational function  $\boldsymbol{f}$ furnishes $\varepsilon-$Differential Privacy if for any neighboring datasets $D$ $\&$ $D^{'}$ and any subset $S$ of all possible outcomes of $\boldsymbol{f}$ obey the following Equation \ref{en:epDP}.
        \end{mydef}
            
        \begin{equation}
        \label{en:epDP}
            \boldsymbol{Pr}[\boldsymbol{f}(D)\; \epsilon \; S] \leq e^{\varepsilon} \times \boldsymbol{Pr}[\boldsymbol{f}(D^{'}) \; \epsilon  \; S ] \; \; \; \; where, \; D\Delta D^{'} = 1
        \end{equation} 

        Here, $\varepsilon$ denotes the privacy budget \& $\Delta$ indicates the maximum difference between the two datasets. However, $\varepsilon-$DP provides a strong privacy guarantee and adds much noise to support the privacy definition by sacrificing the system's utility. It is also called pure differential privacy \cite{beimel2013private}. So, to relax the privacy definition, $(\varepsilon, \delta)-$differential privacy is introduced, which allows quantifying the degree of risk in the privacy guarantee can be tolerated while still ensuring strong overall privacy protection. The formal definition of $(\varepsilon, \delta)-$differential privacy is provided in below section:

        \begin{mydef}
        \label{def:DP2}
        \textbf{$(\varepsilon, \delta)-$Differential Privacy}. 
        A randomized computational function  $\boldsymbol{f}$ ensures $(\varepsilon, \delta)-$Differential Privacy with $\delta$ probability of the privacy guarantee being breached if for any neighboring datasets $D$ $\&$ $D^{'}$ and any subset $S$ of all possible outcomes of $\boldsymbol{f}$ obey the following Equation \ref{en:dDP}.
        \end{mydef} 

        \begin{equation}
        \label{en:dDP}
            \boldsymbol{Pr}[\boldsymbol{f}(D)\; \epsilon \; S] \leq e^{\varepsilon} \times \boldsymbol{Pr}[\boldsymbol{f}(D^{'}) \; \epsilon \; S ] + \delta  \; \; \; \; where, \; D\Delta D^{'} =1 
        \end{equation}

        Usually, the value of $\delta$ is small and $\delta > 0$. If $\delta = 0$, the randomized computational function $\boldsymbol{f}$ provides $\varepsilon-$DP assurance. As $(\varepsilon, \delta)-$DP allows one to choose strict DP for some low probability of privacy breach, it is also called approximate differential privacy \cite{beimel2013private}. This relaxation is advantageous when strict privacy is difficult to achieve without severely compromising utility. However, in the case of multiple training steps of the model, $(\varepsilon, \delta)-$DP cannot guarantee privacy protection as expected. Rényi differential privacy (RDP) can be applied where it guarantees privacy protection with tighter bounds on the privacy budget for multiple training steps. The formal definition of RDP is given below:

        \begin{mydef}
        \label{def:RDP}
        \textbf{$(\alpha, \varepsilon)-$Rényi Differential Privacy} \cite{mironov2017renyi, ning2022eana}. For $1 \, < \,  \alpha \, \leq \, \infty$, a randomized computational function  $\boldsymbol{f}$ ensures $(\alpha, \varepsilon)-$Rényi Differential Privacy if for any neighboring datasets $D$ $\&$ $D^{'}$ with any subset $S$ of all possible outcomes of $\boldsymbol{f}$ and $\varepsilon\, \geq \, 0$ obey the following Equation \ref{en:RDP}.
        \end{mydef}
            
        \begin{equation}
        \label{en:RDP}
            \begin{matrix}
                \mathfrak{D}_\alpha (\boldsymbol{f}(D)||\boldsymbol{f}(D^{'})) \leq \varepsilon & where \; \;D\Delta D^{'} = 1 \\ \\
                \Rightarrow \frac{1}{\alpha -1} log \sum_{s\subset S}(\frac{Pr(\boldsymbol{f}(D) = s)}{Pr(\boldsymbol{f}(D^{'}) = s)})^{\alpha -1} \leq \varepsilon & \\
            \end{matrix}
        \end{equation}
        
        Here, $\mathfrak{D}_\alpha$ denotes Rényi Divergence, which measures the difference between two probability distributions with $\alpha$ order. $s$ is a subset of the output set $S$ from the computational function $\boldsymbol{f}$. It is a generalization of the Kullback-Leibler (KL) divergence, where $\alpha > 1$. If $\alpha \to 1$, Rényi divergence converges to the KL divergence. RDP provides a flexible privacy guarantee using parameters $\alpha$ and $\varepsilon$, resulting in tighter privacy loss bounds. This approach enables the addition of less noise, enhancing the utility of recommendations while maintaining equal or superior privacy protection compared to other DP variations \cite{zheng2023decentralized}. However, the privacy assurance of DP depends on the privacy budget ($\varepsilon$), which refers to the total amount of privacy loss allowed for a particular data analysis task or set of functions. A smaller privacy budget implies more robust privacy because the output distribution is less sensitive to changes in any individual's data by adding excessive noise. In contrast, a larger privacy budget implies weaker privacy. After all, the output distribution can change more significantly when an individual's data is included or excluded. Proper privacy budget selection depends on sensitivity. The sensitivity quantifies how much a single data can influence the outcome of a query or function. It measures the maximum change in the query output when one individual's data is added or removed from the dataset. The formal definition of the sensitivity is as follows:

        \begin{mydef}
        \label{def:sensitivity}
        \textbf{Sensitivity}. For any neighboring datasets $D$ $\&$ $D^{'}$, the sensitivity ($\Delta \boldsymbol{f}$) of a randomized computational function $\boldsymbol{f}$ is defined as:
        \end{mydef}

        \begin{equation}
            \label{en:sensitivity}
            \Delta \boldsymbol{f}_k = max_{D, D^{'}} \left\| \boldsymbol{f}(D) - \boldsymbol{f}(D^{'}) \right\|_k \;\;\; where, \; D\Delta D^{'} = 1
        \end{equation}

        Here, $||.||_k$ denotes $L_k$-norm. Sensitivity controls the noise addition as it determines how much noise needs to be added to achieve DP. The higher the sensitivity, the more noise is required to mask the influence of any single data point, thus ensuring privacy. Usually, Laplace \cite{liu2017differential, yin2020improved, wang2020global, zhang2021graph, neera2021private, yang2021fcmf}, Gaussian \cite{ning2022eana, yang2023fairness, liu2024matrix} \& Exponential \cite{chen2021differentially} mechanisms are commonly applied for noise addition to satisfy the DP. Furthermore, based on RS reliability, DP can be implemented at the system or the user end. Assuming the system is a trusted entity responsible for collecting all user data and ensuring privacy, DP can be implemented within the system. However, if the RS is not trusted, DP can be applied at the user's end, known as Local Differential Privacy (LDP) \cite{wang2020global}. The prime difference between DP \& LDP is that DP considers all users' data as input and requires the output to be indistinguishable, whereas LDP takes into consideration each user's data and randomly perturbs per-user data for downstream tasks.
        
    \end{itemize}
    
\end{itemize}

Table \ref{tab:PPRS_NonCrypto} summarizes all reviewed papers for non-cryptography-based privacy-preserving techniques concerning architecture, approach, targeted problem scenario, performance measurement parameters, and experimental domain with datasets. 

{\fontsize{7.5pt}{7.5pt}\selectfont
\renewcommand{\arraystretch}{0.9}
\begin{longtable}{>{\centering\hspace{0pt}}m{0.086\linewidth}>{\centering\hspace{0pt}}m{0.107\linewidth}>{\centering\hspace{0pt}}m{0.135\linewidth}>{\centering\hspace{0pt}}m{0.195\linewidth}>{\centering\hspace{0pt}}m{0.190\linewidth}>{\centering\arraybackslash\hspace{0pt}}m{0.161\linewidth}}

\caption{Summary of Non-Cryptography-based Privacy-Preserving Recommender System Approaches} \\
\hline 

\textbf{Reviewed Paper} & \textbf{RS Architecture} & \textbf{Privacy Methodology} & \textbf{Considered Problem Scenario} & \textbf{Considered Performance Parameter} & \textbf{Experimental Domain \& Dataset Name} \\

\hline 
\endfirsthead
\multicolumn{5}{c}%
{\tablename\ \thetable\ . Summary of Non-Cryptography-based Privacy-Preserving Recommender System (Cont.)} \\ 
\hline

\textbf{Reviewed Paper} & \textbf{RS Architecture} & \textbf{Privacy Methodology} & \textbf{Considered Problem Scenario} & \textbf{Considered Performance Parameter} & \textbf{Experimental Domain \& Dataset Name} \\
\hline 
\endhead
\hline 
\endfoot
\hline
\endlastfoot

\cite{zhang2019caching} & Decentralized & k-anonymity & Inferring users attribute by the untrusted location service provider & Resistance to eavesdropping attacks, Privacy check \& communication overhead \par{} & POI (Synthetic Dataset) \\ 

\cite{zhao2018illia} & Decentralized & k-anonymity & Data Poisoning Attack & \textbf{1.} Average attack success rate, \textbf{2.} Average cloaking success rate, \textbf{3.} Average processing time \& \textbf{4.} cumulative distribution function \par{} & POI (loc-Gowalla\tablefootnote{\label{note:Loc}https://snap.stanford.edu/data/loc-gowalla.html}) \\ \\

\cite{wei2018improving} & Centralized & k-anonymity, l-diversity \& t -closeness & Lack of assurance of privacy guarantee with high RS accuracy & Information loss, Re-identification probability \& Prediction Accuracy\par{} & Movie (ML@100k)\footref{note:ML} \\ 

\cite{saleem2021parking} & Decentralized & k-anonymity \& DP & Inferring sensitive information & Privacy-Utility trade-off & POI (Parking Dataset) \\ \\

\cite{beigi2020privacy} & Centralized & AL & AIA & Privacy-Utility trade-off \par{} & Movie (ML@100k\footref{note:ML} \\ \\ 

\cite{ganhor2022unlearning} & Centralized & AL & Population Bias & RS accuracy (performance) \& AL attacker accuracy (bias)\par{} & Movie (ML@1M)\footref{note:ML}, \& Music (LFM@2bDB\tablefootnote{\label{note:LMF2B}http://www.cp.jku.at/datasets/LFM-2b})  \\

\cite{yalcin2024novel} & Centralized & Perturbation & Data Leakage & Overhead costs  (Communication, Storage \& Computational cost) \par{}  & Movie (ML@1M\footref{note:ML} \& Netflix\tablefootnote{\label{note:netflix}https://www.kaggle.com/datasets/netflix-inc/netflix-prize-data}) \\ 

\cite{hu2021rap} & Decentralized & Perturbation & Data leakage \& enhancement of RS accuracy & Privacy-Utility trade-off, Communication overhead between public rating server and users \& System scalability & Movie (ML@20M)\footref{note:ML} \& Joke (Jester)\tablefootnote{\label{note:joke}https://grouplens.org/datasets/jester/}  \\ 

\cite{polatidis2017privacy} & Decentralized & Perturbation & Data leakage \& enhancement of RS accuracy & Privacy-Utility trade-off & Movie (ML@100K\footref{note:ML}, MovieTweetings\tablefootnote{\label{note:MT}https://www.kaggle.com/datasets/tunguz/movietweetings}, YahooMovies\tablefootnote{\label{note:YM}https://webscope.sandbox.yahoo.com/catalog.php?datatype=r}, FilmTrust\footref{note:FlimT}) \& Music (YahooAudio)\footref{note:YM} \\ 

\cite{kashani2020feature} & Centralized & k-anonymity \& Perturbation & System's reliability \& implicit information gain from the available data \par{} & Privacy-Utility trade-off & Social Network (Epinion)\footref{note:EP}  \\ 




\cite{base01_2021Gender} & Centralized & Obfuscation & Gender Bias, AIA, Data sparsity, Fairness \& Diversity\par{} & Privacy-Utility trade-off, Fairness \& Diversity & Movie (ML@1M\footref{note:ML}, Flixster) \& Music (LFM@\tablefootnote{\label{note:lm}http://millionsongdataset.com/lastfm/}) \\

\cite{lin2022privacy} & Centralized & Obfuscation & AIA, Gender Bias, Data sparsity \& Long tail items & Privacy-Utility trade-off \& Bias detection\par{}  & Movie (ML@1M)\footref{note:ML} \\

\cite{feng2015can} & Centralized & Obfuscation & AIA & Privacy-Utility trade-off \& Complexity analysis\par{} & Movie (ML@1M\footref{note:ML}, \& Flixster) \\ \\

\cite{salamatian2015managing} & Centralized & Obfuscation & AIA & Privacy-Utility trade-off \par{}  & User Survey \\ \\

\cite{weinsberg2012blurme} & Centralized & Obfuscation & AIA & Privacy-Utility trade-off \par{} & Movie (ML@1M\footref{note:ML}, Flixster) \\ \\



\cite{neera2021private} & Decentralized & Differential Privacy (LDP; $\varepsilon$-DP)\par{} & Decline RS Accuracy & Privacy-Utility trade-off \& Communication Cost & ML@100k\footref{note:ML}, Jester\footref{note:joke} \&  Libimseti \\ \\

\cite{wang2020global} & Centralized \& Decentralized & Differential Privacy (GDP \& LDP; $\varepsilon$-DP)\par{} & User Dependency Graph & Privacy-Utility trade-off & Music (LFM@\footref{note:lm}), Book (Delicious) \& Synthetic dataset \\ \\

\cite{ning2022eana} & Centralized & Differential Privacy $(\alpha, \varepsilon)$-RDP & Secret Sharer Attack \& increased model training time due to applying DP & Model Quality \& privacy, Training speed \par{} & Movie (ML@20M)\footref{note:ML} \& Industry (Industry-100K \& Industry-5M) \\ \\

\cite{liu2024matrix} & Centralized & Differential Privacy ($\varepsilon, \delta$)-DP & Privacy of implicit  data \& data leakage from recommendations & Privacy-Utility trade-off \& model efficiency\par{} & Movie (ML@1M\footref{note:ML}, ML@100k\footref{note:ML} \& FilmTrust\footref{note:FlimT}) \& E-commerce (Amazon\tablefootnote{\label{note:amazon}http://jmcauley.ucsd.edu/data/amazon/links.html})\par{} \\ \\

\cite{mullner2023reuseknn} & Centralized & Differential Privacy ($\varepsilon$-DP) & Inference Attack for fake users generation & Privacy-Utility trade-off, Neighborhood Reuse \& Popularity Bias \par{} & Movie (ML@1M\footref{note:ML}, Douban \& Ciao), Music (LFM@)\footref{note:lm}, \& Books (Goodreads)\par{} \\

\cite{yang2023fairness} & Centralized & Differential Privacy ($\varepsilon, \delta$)-DP & Algorithmic fairness of PPRS & Privacy-Utility trade-off, c-fairness (active \& non-active users) \par{} & E-commerce (Home \& Living, Craft Supplies \& Tool, Grocery \& Gourmet Food, Beauty) \\ 

\cite{chen2021differentially} & Centralized & Differential Privacy ($\varepsilon$-DP) & $k-$nearest neighboring attack \& performance degradation while applying DP in RS \par{} & RS accuracy & Movie (ML@1M\footref{note:ML} \& Netflix\footref{note:netflix}) \\

\cite{zheng2023decentralized} & Decentralized & Differential Privacy (LDP; ($\delta, \varepsilon$)-RDP) & Data leakage through gradient sharing \& RS accuracy downfall through iterative model training with DP & Privacy-Utility trade-off & Movie (Flixster), Music (Bookcrossing) \& POI (Weeplaces) \\ 

\cite{liu2017differential} & Centralized & Perturbation \& Differential Privacy ($\varepsilon$-DP)\par{} & Inference Attack & RS accuracy \& algorithm running time & Movie (ML@100K\footref{note:ML}, Netflix\footref{note:netflix} \& YahooMovies\footref{note:YM}) \\

\cite{zhang2021graph} & Decentralized & Differential Privacy (LDP; $\varepsilon$-DP) & AIA & Privacy-Utility trade-off, Robustness against different AIA & Movie (ML@100k\footref{note:ML}) \\ 

\cite{wang2023improved} & Decentralized & Differential Privacy (LDP; $\varepsilon$-DP) & Privacy of rated items along with rating value by considering AIA & Privacy-Utility trade-off & Movie (ML@10M\footref{note:ML}, FilmTrust\footref{note:FlimT}), Music (YahooMusic)\footref{note:YM} \& Digital Content (Amazon instance video\footref{note:amazon}) \\

\cite{yin2020improved} & Centralized & Differential Privacy ($\varepsilon$-DP) & Enhancement of Accuracy \& Privacy & Privacy-Utility trade-off & Movie (ML@1M)\footref{note:ML} \\ 

\label{tab:PPRS_NonCrypto}
\end{longtable}
}


\subsubsection{\textbf{Advantages and Disadvantages of Cryptography and Non-Cryptography based Methodology.}}

Here, we discuss the benefits and limitations of the PPRS methodologies in Table \ref{tab:MethodPPRS}.  

\begin{table}[h]
\fontsize{7.5pt}{7.5pt}\selectfont
\renewcommand{\arraystretch}{1.2}
\centering
\caption{List of benefits and limitations of the PPRS methods}
\label{tab:MethodPPRS}
\begin{tabular}{>{\centering\hspace{0pt}}m{0.149\linewidth}>{\hspace{0pt}}m{0.307\linewidth}>{\hspace{0pt}}m{0.486\linewidth}}
\hline 

\textbf{PPRS}\par{}\textbf{Methodology} & \multicolumn{1}{>{\centering\hspace{0pt}}m{0.307\linewidth}}{\textbf{Advantages}} & \multicolumn{1}{>{\centering\arraybackslash\hspace{0pt}}m{0.486\linewidth}}{\textbf{Disadvantages}} \\
\hline

Cryptography & \begin{tabular}{@{\labelitemi\hspace{\dimexpr\labelsep+0.5\tabcolsep}}l@{}}Provide secure \& authorized channel \end{tabular}\par{}for data sharing by encrypting it before being processed by the system and prevents unauthorized access and data breaches.
\begin{tabular}{@{\labelitemi\hspace{\dimexpr\labelsep+0.5\tabcolsep}}l@{}}Ensure data privacy by allowing \end{tabular}\par{}computations to be performed on encrypted data without needing to decrypt it first.
\begin{tabular}{@{\labelitemi\hspace{\dimexpr\labelsep+0.5\tabcolsep}}l@{}}Provide strong attack reliance by \end{tabular}\par{}encrypting data and performing secure computations where adversaries try to deduce sensitive information from the data.
\begin{tabular}{@{\labelitemi\hspace{\dimexpr\labelsep+0.5\tabcolsep}}l@{}}Ensure system utility as the advanced ~\end{tabular}\par{}cryptographic methodologies such as HE can be designed to minimize the loss of utility by allowing accurate computations on encrypted data thus ensuring the quality of recommendations remains high. 

& 

\begin{tabular}{@{\labelitemi\hspace{\dimexpr\labelsep+0.5\tabcolsep}}l@{}}Usually, cryptography methods, such as HE, require complex\end{tabular}\par{} mathematical operations, which are computationally intensive. \par
\begin{tabular}{@{\labelitemi\hspace{\dimexpr\labelsep+0.5\tabcolsep}}l@{}}The increased computational requirements can significantly \end{tabular}\par{} slow down the recommendation process, making it impractical for real-time applications \cite{hu2021rap}.\par
\begin{tabular}{@{\labelitemi\hspace{\dimexpr\labelsep+0.5\tabcolsep}}l@{}}RS typically deals with large datasets containing millions of \end{tabular}\par{} users and items where cryptography operations scale poorly with the size of the data, leading to inefficiencies.\par
\begin{tabular}{@{\labelitemi\hspace{\dimexpr\labelsep+0.5\tabcolsep}}l@{}}The storage requirements for encrypted data and the interme$-$\end{tabular}\par{} diate results of cryptography operations can be substantial where users might face limitations to use required service from the RS due to the lack of supported device.\par
\begin{tabular}{@{\labelitemi\hspace{\dimexpr\labelsep+0.5\tabcolsep}}l@{}}Require frequent data exchange between client and system or \end{tabular}\par{} the parties like SMC, which lead to high communication costs.\par
\begin{tabular}{@{\labelitemi\hspace{\dimexpr\labelsep+0.5\tabcolsep}}l@{}}The added computational and communication overhead can \end{tabular}\par{} degrade the user experience \& satisfaction. \par
\begin{tabular}{@{\labelitemi\hspace{\dimexpr\labelsep+0.5\tabcolsep}}l@{}}Some methodology, such as secret sharing, requires constant \end{tabular}\par{} communication with each other by considering online availability of all clients to get expected services from the system \cite{zhang2024hn3s}. \par
\begin{tabular}{@{\labelitemi\hspace{\dimexpr\labelsep+0.5\tabcolsep}}l@{}}As the RS produces approximate results based on encrypted \end{tabular}\par{} data, which may not be as accurate as those obtained from processing raw data. Because the cryptography method encrypts all data, including missing data, which is an expense of algorithm efficiency \cite{ge2021privitem2vec}.  
\\
\hline 

Non-Cryptography 
& 
\begin{tabular}{@{\labelitemi\hspace{\dimexpr\labelsep+0.5\tabcolsep}}l@{}}Non-cryptography methods, such as \end{tabular}\par{} differential privacy \& randomization techniques generally require less computational power~compared to cryptography methods. This characteristic makes them suitable for real-time applications and large-scale datasets.\par
\begin{tabular}{@{\labelitemi\hspace{\dimexpr\labelsep+0.5\tabcolsep}}l@{}}Due to the light computation burden,  ~\end{tabular}\par{}it is easier to understand and implement. For instance, adding Laplace noise to data (DP) or obfuscating data through randomization techniques can be straightforward.\par
\begin{tabular}{@{\labelitemi\hspace{\dimexpr\labelsep+0.5\tabcolsep}}l@{}}As the system is responsible for \end{tabular}\par{}maintaining users' privacy, it reduces the user-end overhead\par
\begin{tabular}{@{\labelitemi\hspace{\dimexpr\labelsep+0.5\tabcolsep}}l@{}}Comparatively easy to integrate into \end{tabular}\par{}existing RS without significant changes to the underlying architecture. 
& 
\begin{tabular}{@{\labelitemi\hspace{\dimexpr\labelsep+0.5\tabcolsep}}l@{}}There is an inherent trade-off between privacy and utility\end{tabular}\par{} (accuracy and relevance) of the recommendations due to adding noises.\par
\begin{tabular}{@{\labelitemi\hspace{\dimexpr\labelsep+0.5\tabcolsep}}l@{}}Properly calibrating the amount of noise to add for privacy \end{tabular}\par{} protection is a very challenging task as insufficient noise can lead to~privacy breaches, while excessive noise can render the data or recommendations useless.\par
\begin{tabular}{@{\labelitemi\hspace{\dimexpr\labelsep+0.5\tabcolsep}}l@{}}Sophisticated adversaries can use background knowledge and\end{tabular}\par{} auxiliary information to infer sensitive data despite non-cryptography protections.\par
\begin{tabular}{@{\labelitemi\hspace{\dimexpr\labelsep+0.5\tabcolsep}}l@{}}Randomization approach, especially random perturbation, is \end{tabular}\par{} foreseeable for the adversary due to adding a random number to create perturb data \cite{kashani2020feature}. Users' data can be retrieved with high accuracy by clustering the perturbed data \cite{xu2017privacy}. \par
\begin{tabular}{@{\labelitemi\hspace{\dimexpr\labelsep+0.5\tabcolsep}}l@{}}Non-cryptographic methods, such as DP, often do not protect\end{tabular}\par{} against inference attacks where an adversary tries to deduce hidden attributes from the available data \cite{chai2022efficient}. Also, these are unable to hide similar neighbours, which creates the scope to conduct a KNN attack \cite{badsha2017privacy}.\par
\begin{tabular}{@{\labelitemi\hspace{\dimexpr\labelsep+0.5\tabcolsep}}l@{}}Anonymization techniques, such as k-anonymity can be~\end{tabular}\par{}vulnerable to de-anonymization attacks.
\\

\hline
\end{tabular}
\end{table}

\subsubsection{\textbf{Hybrid Approach based Methodology}}

Some researchers have proposed a combined approach with cryptography and non-cryptography methodology to ensure utmost privacy preservation and enhancement of recommendation accuracy. The author of \cite{zigomitros2016practical} proposed a practical k-anonymity implementation with hashing encryption to mitigate each technique's limitation and retrieve the cumulative benefits of the approaches. In \cite{zigomitros2016practical}, the client used the local hash function to encrypt the data and send the data to the system. The system added random noise by applying k-anonymity to prevent data leakage from the recommendations. Hu et al. \cite{hu2023differentially} applied DP with a local hash function to protect privacy in FedRec by considering the drawback of the hash function. Wang et al. \cite{wang2021cryptorec} also proposed a hybrid privacy preservation approach by considering three problems that are user attribute inference, data leakage from recommendation lists, and privacy breach from knowledge sharing between datasets with the same items profiles. To mitigate data leakage and transfer knowledge, the authors consider $\varepsilon$-differential privacy. However, the DP also leakages data with iterative model training. Consequently, the authors applied homomorphic encryption to counter user attribute inference and DP weakness. Conversely, the authors of \cite{chai2022efficient} also used the same hybrid approach in the federated environment by considering inference attacks. They applied DP to disguise users' data to protect against data leakage and HE to protect the intermediate results of the model during the training stage. Yang et al. \cite{yang2021fcmf} followed and proposed the same hybrid approach by taking into account privacy breaches that might initiated due to collaborating different formats of user feedback data in FedRec architecture. Table \ref{tab: PPRS_hybrid} explains each reviewed paper regarding the proposed approach's privacy technique, problem scenarios, and experimental parameters with the datasets.  

\begin{table}[h]
\fontsize{7.5pt}{7.5pt}\selectfont
\renewcommand{\arraystretch}{0.9}
\centering
\caption{Summary of Hybrid Privacy-Preserving Recommender System Approaches}
\begin{tabular}{>{\centering\hspace{0pt}}m{0.10\linewidth}>{\centering\hspace{0pt}}m{0.156\linewidth}>{\centering\hspace{0pt}}m{0.250\linewidth}>{\centering\hspace{0pt}}m{0.2\linewidth}>{\centering\arraybackslash\hspace{0pt}}m{0.20\linewidth}}
\toprule
\textbf{Reviewed Paper} & \textbf{Privacy Methodology} & \textbf{Considered Problem Scenario} & \textbf{Considered Performance Parameter} & \textbf{Experimental Domain and Datasets Name} \\
\midrule

\cite{chai2022efficient} & HE \& Differential \par{}Privacy ($\varepsilon$-DP) & Data leakage \& intermediate training result leakage & Privacy-Utility trade-off, \& efficiency (communication time)\par{} & Movie (ML@100k\footref{note:ML}) \&  POI (users check-in) \\ 


\cite{yang2021fcmf} & HE \& Differential Privacy ($\varepsilon$-DP) & Data leakage from the combination of heterogeneous users feedback and intermediate model training result leakage & Privacy-Utility trade-off & Movie (ML@100k, ML@1M, ML@10M\footref{note:ML} \& Netflix\footref{note:netflix}) \\ 

\cite{zigomitros2016practical} & Hashing \& K-anonymity & Data leakage from user \& RS exposure & Privacy-Utility trade-off, Communication \& Computational Cost & Movie (ML@1M)\footref{note:ML} \\ 


\cite{hu2023differentially} & Hashing (locality sensitive hashing) \& Differential Privacy (LDP; $\varepsilon$-DP) & Actual data leakage \& potential data leakage from exposure & Privacy-Utility trade-off & Image (MNIST\tablefootnote{\label{note:image}http://yann.lecun.com/exdb/mnist/}), Document (Bag of Words\tablefootnote{\label{note:word}https://archive.ics.uci.edu/dataset/164/bag+of+words}) \& Joke (Jester)\footref{note:joke} \\ \\

\cite{wang2021cryptorec} & HE \& Differential Privacy ($\varepsilon$-DP) & Users data leakage to the system \&  from the recommendation lists & Privacy-Utility trade-off, computational cost \& knowledge transfer & Movie (ML@1M\footref{note:ML}, Netflix\footref{note:netflix} \& YahooMovies\footref{note:YM} ) \\ 

\bottomrule
\end{tabular}
\label{tab:PPRS_hybrid}
\end{table}

\section{Discussion and Future Directions}
\label{sec:fd}

In recent years, academia and industry have focused more on beyond accuracy aspects of recommender systems, emphasizing their importance. For example, Slokom et al. \cite{base01_2021Gender} explained that gender bias in data harms fairness in recommendations and creates opportunities for attribute inference attacks, threatening users' privacy. They used obfuscation techniques to address this privacy issue by mixing users' rated data in a way that an adversary can not infer users' gender from their ratings. However, this approach might introduce popularity bias in the data, affecting p-fairness. Subsequently, Lin et al. \cite{lin2022privacy} addressed the same issue while accounting for popularity bias by incorporating long-tail (unpopular) items. Despite these efforts, these privacy-preserving strategies still fell short of providing comprehensive privacy guarantees. Yang et al. \cite{yang2023fairness} noted that formal privacy guarantees couldn't ensure fairness, leading to biased recommendations. They proposed a solution that balanced fairness and privacy, albeit at the cost of accuracy. Additionally, existing research often focuses on isolated aspects and proposes solutions without considering the interconnections between different aspects. In the next section, we list future research directions based on the overall limitations of current studies.

\subsection{Bias Detection, Exploration and Mitigation Frameworks with Fairness-Aware}

Addressing bias in the recommender system is crucial for ensuring fairness and equity in the recommendations provided to users while also considering item fairness. Bias can occur in any component of the RS, including data, algorithms, or user actions. Developing methods to identify, measure, and mitigate biases in the recommendation process, such as data collection, algorithm design, and output generation, is a promising research direction. This could involve creating fairness-aware algorithms that provide equitable recommendations across diverse user groups, ensuring that no particular demographic is unfairly advantaged or disadvantaged. Techniques such as debiasing data preprocessing, fairness constraints in model training, and post-processing adjustments to recommendations can be explored. Additionally, research could focus on developing metrics and frameworks for evaluating the fairness of recommender systems, enabling an accountable assessment of their performance. RS can be more inclusive and just by systematically addressing these biases, fostering trust and satisfaction among all users and inspiring future researchers to contribute to this critical and evolving field.

As data bias initiates and exhilarates other biases in any RS components, it could be a hot research topic to integrate heterogeneous data (explicit and implicit data together) to detect bias in the data and pre-process the training data by considering diversity and imposing fairness. In the real world, bias is not static; it is dynamic. Moreover, biases are interrelated, and a bias can intensify other biases over time. For example, ML-based recommender systems analyze data and capture patterns for subsequent processing. If the model’s input data is biased, it will lead to algorithmic bias because data is the primary cause of algorithmic bias and can intensify its degree. The effect of data bias on algorithmic bias is called \textbf{compound imbalances} \cite{de2019bias, hellman2018indirect}. Furthermore, one bias can introduce others; for instance, selection bias might lead to position and popularity biases in the system. There is also a correlation between popularity and gender biases \cite{wang2021user, ekstrand2018all}, and personality bias is also positively correlated with position bias. Exploring the evolution of bias and analyzing its dynamic nature's impact on RS will be insightful topics for future research.

Moreover, popularity bias can act as a double-edged sword \cite{zhang2021causal, zhao2022popularity}. The algorithm initiates popularity bias and reflects the item’s quality within the dataset's prevalent skew. An appropriate proportion of popularity bias in the system can improve performance and user satisfaction \cite{zhao2022popularity}. Investigating whether other biases have a double-edged sword nature is another interesting research topic that could enhance system performance by controlling the existence of biases with benign effects.

\subsection{Fairness-Accuracy-Privacy Trade-off}

The trade-off between the trio (fairness, accuracy, and privacy) is crucial in recommendation systems. In scenarios where a non-cryptographic approach is applied to ensure privacy, there is often a resulting decrease in accuracy and compromise of fairness. Additionally, ensuring fairness in privacy-preserving recommender systems (PPRS) may not provide equal recommendation accuracy for users who frequently use the system. This situation motivates researchers to investigate further: 

\begin{enumerate}
    \item identify specific unfairness issues,
    \item determine the consequences of these unfairness issues at every stage of the RS, and
    \item defines an appropriate privacy framework with fair constraints while maintaining balanced recommendation accuracy.
\end{enumerate}

\subsection{Evaluation Frameworks}

A comprehensive evaluation framework is essential for advancing the development of secure, private, unbiased, and fair recommender systems. Developing standardized, multi-dimensional evaluation frameworks that rigorously assess RS across these critical areas is another interesting and valuable topic of research. Such a framework should incorporate metrics for detecting and mitigating various threats and biases to ensure robust privacy protections and promote fairness across diverse user groups. By integrating these aspects into a unified evaluation framework, researchers and practitioners can systematically benchmark and improve the performance of RS, driving progress toward more reliable and trustworthy recommendations. This holistic approach will clarify how different strategies impact system integrity, user privacy, and overall fairness, guiding the development of next-generation RS.

\subsection{Cross-Domain RS}

Nowadays, cross-domain recommender systems are becoming a promising research area, addressing the limitations of current systems. This method applies user data and preferences from multiple domains (e.g., movies, books, music) to improve recommendation accuracy and diversity \cite{wang2021cross}. By integrating data from various domains, these systems can solve issues like data sparsity and cold-start problems, offering more personalized suggestions \cite{wang2021cross}. However, this approach also brings new challenges in privacy and data security. Combining data from different sources increases the risk of exposing sensitive user information and biased knowledge transfer. Future research should develop advanced algorithms to handle cross-domain data efficiently and securely, ensuring strong privacy protections while maintaining high-quality recommendations. This direction has excellent potential to create more user-centric systems that cater to diverse needs and preferences.

\section{Conclusion}
\label{sec:con}

This survey has brought to light the multifaceted challenges faced by recommender systems, such as fairness, bias, threats, and privacy. As recommender systems increasingly shape user experiences and decision-making across various platforms, addressing these challenges becomes more pressing. Ensuring fairness is not just a goal but necessary to provide equitable recommendations for all users. Mitigating bias is equally crucial to prevent the reinforcement of existing prejudices. Protecting against threats is vital to maintain the integrity and reliability of these systems. Furthermore, privacy concerns must be meticulously addressed to safeguard user data and maintain trust. In each section, we summarize a collection of influential research work, highlighting the main concepts with our insights. Moreover, we also extend probable promising future research direction by focusing on developing advanced methodologies and robust frameworks that enhance the accuracy and efficiency of recommender systems and ensure they operate in a fair, secure, and privacy-preserving manner. Although a large proportion of research work and techniques are proposed each year, we hope this study can aid readers and developers in quickly understanding the fundamental aspects and importance of accuracy beyond parameters of the RS for real-life applications.

\bibliographystyle{unsrt}

\end{document}